\documentclass[%
 reprint,
 amsmath,amssymb,
 aps,
]{elsarticle}
\usepackage{comment}
\usepackage{balance}
\usepackage{crimson}
\usepackage[margin=.8in]{geometry}
\usepackage{blindtext}
\usepackage[parfill]{parskip}  
\usepackage{graphicx}
\usepackage{amssymb}
\usepackage{empheq}
\usepackage{braket}
\usepackage{tikz} 
\usetikzlibrary{shapes,arrows,positioning,automata,backgrounds,calc,er,patterns}
\usepackage{tikz-feynman}
\usepackage{slashed}
\usepackage{epstopdf}
\usepackage{mathtools}
\usepackage{wrapfig}

\newcommand{\be}{\begin{equation}}
\newcommand{\ee}{\end{equation}}
\newcommand{\ba}{\begin{align}}
\newcommand{\ea}{\end{align}}
\usepackage[bottom]{footmisc}

\usepackage{subcaption}

\begin{document}
\title{Explicit computation of jet functions in coordinate space}
\author{Alexandre Salas-Bern\'ardez\textsuperscript{$\dagger$}\\
\textsuperscript{$\dagger$} Universidad Complutense de Madrid, Departamento de F\'isica Te\'orica and IPARCOS, Plaza de las Ciencias s/n, 28040, Madrid.}

\begin{abstract}
\centering
\begin{minipage}{\dimexpr\paperwidth-6.85cm}

  $\;\;$ I review the main results leading to Factorization of QCD amplitudes in momentum space and, in view of the analogue results in coordinate space, the one-loop jet function in coordinate space is computed and Landau's equations for the Abelian radiative corrections to it are studied. Furthermore, two of these radiative corrections are reduced in quadrature.
\end{minipage}
\end{abstract}
\maketitle

\section{Introduction}

In Quantum Field Theory (QFT), Green's functions are the mathematical entities used to relate the theoretical framework to physical processes. In principle, if all the Green's functions of a particular theory are known, then the theory is solved completely (since these functions are introduced to solve the quantum equations of motion for the fields). Usually, the computation of Green's functions is performed in momentum space due to mathematical simplicity. Furthermore, scattering and collider experiments, which are among the main methods and tools to unravel the structure of elementary particles, are formulated and studied in momentum space. Here, the requirement of unitarity (forcing conservation of probabilities) naturally arises as the optical theorem and helps identifying resonances or bound states that the theory possesses. Moreover, huge efforts were made at the end of last century to justify the consistent use of perturbation theory in Quantum Chromodynamics (QCD). These resulted in the factorization theorems for amplitudes in momentum space, which, by means of Resummation, ensured the convergence of the perturbation series in the weak coupling regime (see \cite{Collins} or \cite{Stermannotes}). On the other hand, analogue results in coordinate space only appeared in recent years \cite{ErdoganCS}, signaling the little attention that the coordinate space description has received in high-energy physics research. Seeking new perspectives and interpretations, I will review several aspects of QCD Amplitudes from the coordinate space viewpoint. \par

This paper is organized as follows: Section \ref{momfact} is devoted to offer a concise introduction to the topics regarding Leading Power (LP) factorization of QCD amplitudes in the perturbative regime. Here Landau's equations are introduced as a condition for having an unavoidable singularity in a momentum space Feynman amplitude. These equations for the one-loop quark electromagnetic form factor, together with the Coleman-Norton picture, already elucidate the all-order structure of divergences in the $\gamma qq$ vertex, \textit{i.e.} a hard function, two jets and a soft function which will be explained. After power counting the possible singularities, factorization of the vertex is possible. The momentum space factorization results were translated to coordinate space amplitudes in O. Erdogan's work \cite{ErdoganCS} and are summarized in Section \ref{coordfact} . Using these results I compute in Section \ref{computations} the one-loop jet function in coordinate space. \\

In recent years, within the community of QCD phenomenologists, interest has grown in the all-order structure of the so called subleading regions. These kinematical regions can enhance non-analyticites in certain observables such as cross sections (see f.e. the first equation in \cite{Bonocore:2015esa}), usually related to the emission of soft and collinear gluons. Explicitly, at Next-to-Leading Power (NLP) in the soft and collinear expansion, one treats with the so-called ``radiative jets'' (introduced for the first time in \cite{DelDuca:1990gz}). 
Given that the study of radiative jet functions in coordinate space could give interesting insights on the still incomplete NLP all-order factorization of QCD (many recent developments show how rich and complicated is the structure of these subleading regions \cite{Gervais:2017yxv,Beneke:2017ztn,Moult:2019mog,Laenen:2020nrt,Liu:2021mac}), in Section \ref{radiativesection} I will reduce to Feynman parameter integrals two contributions to the radiative one-loop jet function in coordinate space. I also analyze Landau's equations for all contributions to the abelian one-loop radiative jet function in coordinate space.

\section{Landau's equations and Leading Power Factorization}\label{momfact}
Amplitudes in QFT encode the probabilities of certain processes contributing to a particular outcome of a collider experiment with known initial conditions. Usually, amplitudes present singularities (which contain all the relevant information of it since finite terms depend on the regularization scheme), and the characterization of these is important to identify their associated kinematical configurations. Each contributing process to an amplitude is represented by a Feynman diagram or graph. A general momentum space Feynman diagram in $d\equiv4-2\epsilon$ spacetime dimensions $G(p_1,...,p_n)$ is given by (leaving out coupling constants for now)
\begin{equation}
G(p_1,...,p_n)=\prod_{i=1}^L\int \frac{d^dk_i}{(2\pi)^d} \prod_{j=1}^I\frac{N(\{k_i\},\{p_r\})}{l^2_j+m_j^2-i\eta}\;,\label{graph}
\end{equation}
where the $\{p_r\}$ are the external momenta, $\{k_i\}$ the loop momenta, $\{l_j\}$ the line momenta (which are linear combinations of the loop and external momenta) and $m_j$ is the current mass for the particle propagating in line $j$. $L$ is the number of loops,  $N(\{k_i\},\{p_r\})$ is an arbitrary polynomial in the momenta and $I$ the number of internal lines. Introducing Feynman's parametrization in eq. (\ref{graph}) we obtain
\begin{align}
G=&(I-1)!\prod_{i=1}^L\int d^dk_i\prod_{j=1}^I\int_0^1d\alpha_j\delta(1-\sum_{j=1}^I\alpha_j)\frac{N( \{k\},\{p\})}{(D(\{\alpha\}, \{k\},\{p\}))^I}.\label{eq:land1}
\end{align}
The graph $G$ will be infrared (IR) divergent if its denominator vanishes unavoidably, $D\equiv\sum_{j=1}^I\alpha_j(l_j^2+m_j^2)=0$. To see what unavoidably means notice that $D$ is a quadratic function in the momenta and therefore it has no more than two zeroes in any momentum component $l_i^\mu$. If these zeroes are at real values of $l_i^\mu$ we will encounter singularities of the integrand in eq. (\ref{eq:land1}). However, if the contour of integration can be deformed in each of the integration variables' complex plane, then, by virtue of Cauchy's theorem, the integral will be well defined and will not present singularities. \par 
There can be two cases where deforming the integration contour will not be possible:
\begin{itemize}
\item If the two zeroes in $k^\mu_i$ merge at the same point and ``pinch'' the contour of integration. This condition amounts to, 
\begin{equation}
\frac{\partial D(p_r,k_i,\alpha_j)}{\partial k_i^\mu}\Big|_{D=0}=0\;,\label{EQ:111}
\end{equation}
which by the definition of $D$ means
\begin{equation}
\sum_{j=1}\alpha_j\frac{\partial l^2_j(p_r,k_i)}{\partial k_i^\mu}=\sum_{j\in loop \;i}\alpha_jl_j^\mu\epsilon_{j,i}=0\;.
\end{equation}
Where the incidence matrix element $\epsilon_{j,i}$ is $+1$ if the line momentum $l_j$ in the loop $i$ flows in the same direction as the loop momentum $k_i$, $-1$ if in the opposite direction and zero otherwise. The sum runs over all the edges in loop $i$.
\item If the singularity is at the endpoints of the contour of integration then we will not be able to apply Cauchy's theorem: we cannot modify the endpoints of integration without affecting the result of the integral. Since $k^\mu_i\in \mathbb{R}$ this type of singularities corresponds to Ultraviolet (UV) divergences and are taken care by renormalization. For $\alpha_j$ integrations these singularities are important when $\alpha_j=0$ or, if $D$ does not depend on $\alpha_j$ (meaning that $l_j^2=-m^2_j$, \textit{i.e.} the line is on-shell). Either one of these two conditions on all of the $\alpha$s has to apply in order to have an unavoidable singularity.
\end{itemize}
Hence, we can condense the necessary conditions for unavoidable divergences in Landau's equations
\begin{empheq}[left=\empheqlbrace]{align}
\sum_{j\in loop \;i}\alpha_jl_j^\mu\epsilon_{j,i}=0 &\;\;\;\; \forall \mu,i \nonumber\\
  \alpha_j(l_j^2+m_j^2)=0 & \;\;\;\;\forall j\;\;\;.   \label{eq:Landau}
\end{empheq}
The proof that these conditions are necessary but not sufficient can be found in \cite{Tod} (pg. 98). To verify that the solutions are indeed unavoidable singularities we have to resort to the method of power counting,  i.e. count the divergence of the integrand and the volume of integration, to see if they give indeed a power or logarithmic divergence when approaching certain divergent kinematical configurations. Furthermore, in the next section we will see that these configurations must be allowed by classical free-particle propagation of the internal lines.

\subsection{Coleman-Norton picture in momentum space}
Finding solutions to Landau's equations (\ref{eq:Landau}) by hand is not an easy task, especially for higher-order diagrams. However, owing to Coleman and Norton (C-N) \cite{ColNor}, there is a much easier and intuitive procedure to solve the equations.\\

Recall that, in a solution to Landau's equations, for off-shell lines we have $\alpha_j=0$ while for an on-shell internal line we will have that $\alpha_j\neq0$ and ${\partial D}/{\partial k_i^\mu}=0$. Now, if we identify the products $\alpha_j l_j$ for each on-shell line with a spacetime vector (introducing a parameter $\lambda$)
\begin{equation}
 \Delta x_j^\mu\equiv\lambda\alpha_jl_j^\mu\;,
\end{equation}
and $\lambda\alpha_j=\Delta x_j^0/l_j^0$ as the Lorentz invariant ratio of the time component of $\Delta x_j^0$ to the energy $l_j^0$, then we have that
\begin{equation}
 \Delta x_j^\mu=\Delta x_j^0v_j^\mu\;,
\end{equation}
with the four-velocity $v_i^\mu=(1,\vec{l}_j/l_j^0)$.  Notice that the parameter $\lambda$ has dimensions of length squared to keep the Feynman parameter dimensionless and its introduction has a subtle meaning. For collinear divergences it can be set to unity but it is necessary in the soft case to keep the displacement $\Delta x^\mu$ finite even when all components of $l^\mu$ go to zero. Since soft gluons have almost infinite wavelength, it is natural to think that they will have a finite displacement as classical particles.

 Summarizing, $\Delta x_j^\mu$ may be thought of as a four-vector describing the free propagation of a classical on-shell particle with momentum $l_j$. In this way, Landau's Equations become

\begin{empheq}[left=\empheqlbrace]{align}
\sum_{j\in loop \;i}\Delta x_j^\mu\epsilon_{j,i}=0 &\;\;\;\; \text{if}\;\;l^2_j=-m_j^2 \nonumber\\
  \Delta x_j^\mu=0\;\;\;& \;\;\;\;\text{if}\;\;l^2_j\neq-m_j^2\;.   \label{eq:Landau2}
\end{empheq}

This means that the ``pinch'' condition for on-shell lines amounts to the requirement that every loop made out of these lines is a closed classical path and that off-shell lines are shrunk to a point (i.e. they do not propagate a finite distance). We hence construct all the possible displacement configurations inside the loops (shrinking or taking the lines as soft) and then decide if these are allowed by the C-N picture. We will regard these as \textbf{reduced diagrams}. To illustrate this method we now turn to an example.
\subsubsection{One-loop Quark Electromagnetic Form Factor reduced diagrams}

Let's now work on an illustrative example of the application of the Coleman-Norton trick to find possible ``pinched'' singularities. In Feynman gauge, the only relevant diagram contributing to the QCD one loop correction to the quark-quark-photon ($qq\gamma$) vertex is
\begin{equation}
\centering
\begin{tikzpicture}
\begin{feynman}
\vertex (a);
\vertex [below=1.3 cm of a] (b);
\vertex [below=0.1cm of b] (H){};
\vertex [right=2.5cm of b] (Hh){$\equiv-ie\,\Gamma_{(1)}^\mu(p_+,p_-)$};
\vertex [below right=1.4cm of b] (k);
\vertex [below right=1cm of k] (t);
\vertex [below left=1.4cm of b] (f1);
\vertex [below left=1cm of f1] (c);

\diagram* {
(a) -- [photon] (b);  (k)-- [gluon] (f1);
(b)-- [] (k);
(k)-- [fermion] (b);
(k)-- [momentum=$p_-$] (t);
(t)-- [fermion] (k);
(f1)-- [fermion] (c);
(c)-- [momentum=$-p_+$] (f1);
(f1) -- [] (b);
(b) -- [fermion] (f1)};
\end{feynman}
\end{tikzpicture}\nonumber\;.
\end{equation}
In physical gauges, there are extra contributions to the form factor coming from pinch surfaces in self energy diagrams.
There are several phenomenological uses of this vertex, some examples can be found in \cite{Bicudo:1998qb,Brodsky:1983vf}.\\

Which reduced diagrams (i.e. diagrams with shrinked lines or with soft lines) of the diagram above are allowed by classical free propagation between vertices?
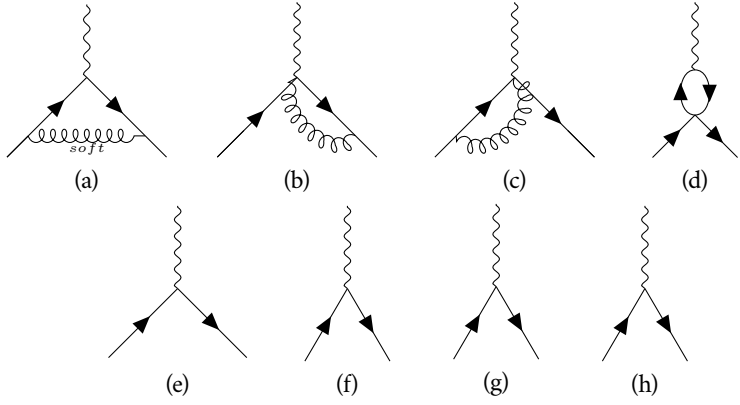
\begin{figure}[ht!]
\centering
\begin{tikzpicture}
\begin{feynman}
\vertex (a);
\vertex [below=0.97 cm of a] (b);
\vertex [below=1.1cm of b] (H){(a)};
\vertex [below=0.75cm of b] (H){\tiny$soft$};
\vertex [below right=1.1cm of b] (k);
\vertex [below right=0.4cm of k] (t);
\vertex [below left=1.1cm of b] (f1);
\vertex [below left=0.4cm of f1] (c);

\diagram* {
(a) -- [photon] (b);  (f1)-- [gluon] (k);
(b)-- [fermion] (k);
(k)-- [] (t);
(c)-- [] (f1);
(c)-- [] (f1);
(f1) -- [fermion] (b);

};
\end{feynman}
\end{tikzpicture}\;\;\;\;\;\;
\begin{tikzpicture}
\begin{feynman}
\vertex (a);
\vertex [below=1 cm of a] (b);
\vertex [below=1.1cm of b] (H){(b)};
\vertex [below right=1.1cm of b] (k);
\vertex [below right=0.1cm of b] (hh);
\vertex [below right=0.4cm of k] (t);
\vertex [below left=1cm of b] (f1);
\vertex [below left=0.5cm of f1] (c);

\diagram* {
(a) -- [photon] (b);  (k)-- [gluon,half left,looseness=1.2] (b);
(b)-- [fermion] (k);
(k)--  (t);
(c)--  (f1);
(c) -- [fermion] (b);

};
\end{feynman}
\end{tikzpicture}\;\;\;\;\;\;\;
\begin{tikzpicture}
\begin{feynman}
\vertex (a);
\vertex [below=1 cm of a] (b);
\vertex [below=1.1cm of b] (H){(c)};
\vertex [below right=1cm of b] (k);
\vertex [below right=0.1cm of b] (hh);
\vertex [below right=0.5cm of k] (t);
\vertex [below left=1.1cm of b] (f1);
\vertex [below left=0.4cm of f1] (c);

\diagram* {
(a) -- [photon] (b);  (b)-- [gluon,half left,looseness=1.4] (f1);
(b)-- [fermion] (t);
(k)--  (t);
(c)--  (f1);
(f1) -- [fermion] (b);

};
\end{feynman}
\end{tikzpicture}\;\;\;\;\;\;\;
\begin{tikzpicture}
\begin{feynman}
\vertex (a);
\vertex [below=0.9 cm of a] (b);
\vertex [below=1.2cm of b] (H){(d)};
\vertex [below=0.6 cm of b] (k);
\vertex [below right=0.8 cm of k] (t);
\vertex [below left=0.8 cm of k] (s);

\diagram* {
(a) -- [photon] (b)  ;
(b) -- [fermion,half left,looseness=1.1] (k) ; 
(k) -- [fermion,half left,looseness=1.1] (b)  ;
(k) -- [fermion] (t);  
(s) -- [fermion] (k);  

};
\end{feynman}
\end{tikzpicture}\;\;\;\;\;\;\;\\
\begin{tikzpicture}
\begin{feynman}
\vertex (a);
\vertex [below=1.1 cm of a] (b);
\vertex [below=0.3 cm of b] (b1);
\vertex [below=1.cm of b] (H){(e)};
\vertex [below right=1.3 cm of b] (t);
\vertex [below left=1.3 cm of b] (s);

\diagram* {
(a) -- [photon] (b) ; 
(b) -- [fermion] (t) ; 
(b) -- [gluon,out=-110, in=-70, loop, min distance=0.9cm] (b);
(s) -- [fermion] (b) ; 

}; 
\end{feynman}
\end{tikzpicture}\;\;\;\;\;\;\;
\begin{tikzpicture}
\begin{feynman}
\vertex (a);
\vertex [below=1.1 cm of a] (b);
\vertex [below=0.3 cm of b] (b1);
\vertex [below=1.cm of b] (H){(f)};
\vertex [below right=0.8 cm of k] (t);
\vertex [below left=0.8 cm of k] (s);

\diagram* {
(a) -- [photon] (b) ; 
(b) -- [fermion] (t)  ;
(b) -- [fermion,out=135, in=-135, loop, min distance=1cm] (b);
(s) -- [fermion] (b)  ;

}; 
\end{feynman}
\end{tikzpicture}
\;\;\;\;\;\;\;
\begin{tikzpicture}
\begin{feynman}
\vertex (a);
\vertex [below=1.1 cm of a] (b);
\vertex [below=0.3 cm of b] (b1);
\vertex [below=1.cm of b] (H){(g)};
\vertex [below right=0.8 cm of k] (t);
\vertex [below left=0.8 cm of k] (s);

\diagram* {
(a) -- [photon] (b);  
(b) -- [fermion] (t) ; 
(b) -- [fermion,out=45, in=-45, loop, min distance=1cm] (b);
(s) -- [fermion] (b)  ;

}; 
\end{feynman}
\end{tikzpicture}
\;\;\;\;\;\;\;
\begin{tikzpicture}
\begin{feynman}
\vertex (a);
\vertex [below=1.1 cm of a] (b);
\vertex [below=1.cm of b] (H){(h)};
\vertex [below right=0.8 cm of k] (t);
\vertex [below left=0.8 cm of k] (s);

\diagram* {
(a) -- [photon] (b) ; 
(b) -- [fermion] (t)  ;
(s) -- [fermion] (b)  ;

};
\end{feynman}
\end{tikzpicture}
\caption{Reduced diagrams for the one-loop QCD correction to the $qq\gamma$ vertex.}
\label{fig:reddiag}
\end{figure}

 We can see all the reduced diagrams in Fig. \ref{fig:reddiag}, each corresponding to an \textit{a priori} solution to Landau's equations. Thanks to the Coleman-Norton Trick we can immediately rule out the diagrams (e)-(g) since a particle that leaves a vertex cannot come back to the same vertex in classical free flight. Diagram (d) is also ruled out since the photon is taken to be off-shell and hence the particles leaving the vertex must have different directions and can never meet again in free-flight (here we assume that the quarks are massless). Diagram (h) corresponds to having all the internal particles off-shell, meaning that this solution to Landau's equations represents an UV divergence. In this way we are left only with diagrams (a)-(c) as possible candidates to IR soft and collinear divergences.

Already with this example it is possible to start picturing the all-order structure of the IR and UV pinched singularities for the electromagnetic quark form factor. In Fig. 1, (a) represents the first order contribution to what will be called the Soft function,  (b) and (c) are each the first order contribution to the two jet functions, and (h) corresponds to the Hard function encoding the UV singularities. 
\subsection{Power-counting and Factorization}
\begin{figure}[h!]
\centering
\begin{tikzpicture}
\shadedraw[inner color=white,outer color=gray,draw=black,rotate=30.7]  (3,0.005) ellipse (40pt and 10pt);
\shadedraw[inner color=white,outer color=gray,draw=black,rotate=-30.7]  (3,-0.005) ellipse (40pt and 10pt);
\shadedraw[inner color=white,outer color=gray,draw=black,rotate=-30]  (0,0) ellipse (10pt and 10pt);
\shadedraw[inner color=white,outer color=gray,draw=black]  (3,0) ellipse (15pt and 15pt);
\draw [thick](0.3,-0.2) -- (4.5,-2.7);
\draw [thick](0.3,0.2) -- (4.5,+2.7);
\draw (4.5,+2.9) node {$p_+$};
\draw (4.5,-2.9) node {$p_-$};
\draw (2.6,-2.4) node {$J_{-}$};
\draw (2.6,2.4) node {$J_{+}$};
\draw (0,0) node {$H$};
\draw (3,0) node {$S$};
\draw [decoration={aspect=0, segment length=1.6mm, amplitude=0.7mm,coil},decorate] (-0.35,0)   -- (-1.5,0)node {\tiny$|$};

 \draw [thick, dotted] (0,0.35) .. controls (0,1.2) and (1.2,1.5) .. (1.7,1.35);
  \draw [thick, dotted] (0,-0.35) .. controls (0,-1.2) and (1.2,-1.5) .. (1.7,-1.35);
 \draw [thick, dotted] (0.35,0) -- (2.47,0);
 \draw [thick, dotted] (2.7,-1.2) -- (3,-0.53);
  \draw [thick, dotted] (2.7,1.2) -- (3,0.53);

\end{tikzpicture}
\caption{Representation of the hard $(H)$, the soft $(S)$, and jet $J_{(\pm)}$ pinch surfaces for the quark electromagnetic vertex at all orders before power counting (see below). The dotted lines represent all the possible lines (fermionic or gluonic) connecting different pinch surfaces. }\label{pinch}
\end{figure}
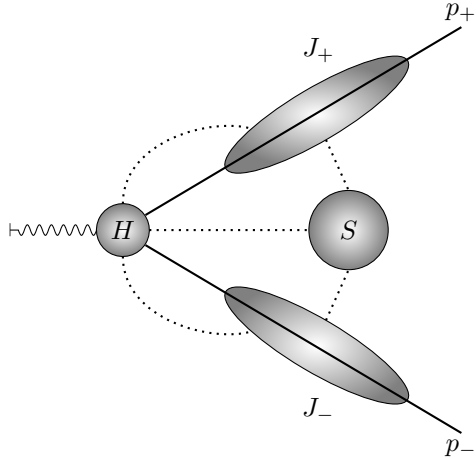

Thanks to the picture of reduced diagrams, we can study the IR and UV divergences in the electromagnetic form factor to \textbf{all orders in perturbation theory} without explicitly considering Landau's equations. It turns out that the structure of singularities we have already studied at one-loop is also present at higher orders. The possible reduced diagrams associated with pinch surfaces are all of the form shown in Figure \ref{pinch}.

The reduced diagram in Fig. \ref{pinch} corresponds to physical processes in which the photon decays into two jets $J_+$ and $J_-$ each with the total momenta of the two final state particles, $p_1$ and $p_2$. Between these two jets the only interaction is via zero-momentum soft particles, labeled $S$. This is due to the fact that once the jets are formed, they travel in different directions at the speed of light and hence no finite momentum transfer can occur between the two. These interactions result on possible phase shifts on the final states due to the inter-quark potential (see \cite{Eric1}). Higher order off-shell, short-distance contributions coming from shrunk lines are encoded in the subdiagram $H$. The full derivation of this characterization of general pinch surfaces to all orders in perturbation theory is presented in \cite{Sterman1}. The term pinch surface is introduced to illustrate the fact that the singularity configurations constrain loop momenta, defining a hypersurface in the $(\{\alpha\},\{k\})$ space where the singularity pinches the contour of integration and makes the singularity unavoidable. 
\par

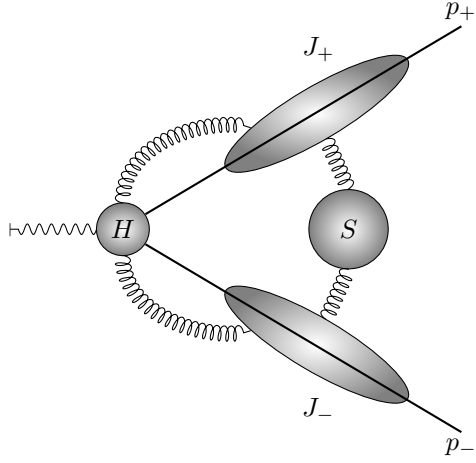
\begin{figure}[ht!]
\centering
\begin{tikzpicture}
\shadedraw[inner color=white,outer color=gray,draw=black,rotate=30.7]  (3,0.005) ellipse (40pt and 10pt);
\shadedraw[inner color=white,outer color=gray,draw=black,rotate=-30.7]  (3,-0.005) ellipse (40pt and 10pt);
\shadedraw[inner color=white,outer color=gray,draw=black,rotate=-30]  (0,0) ellipse (10pt and 10pt);
\shadedraw[inner color=white,outer color=gray,draw=black]  (3,0) ellipse (15pt and 15pt);
\draw [thick](0.3,-0.2) -- (4.5,-2.7);
\draw [thick](0.3,0.2) -- (4.5,+2.7);
\draw (4.5,+2.9) node {$p_+$};
\draw (4.5,-2.9) node {$p_-$};
\draw (2.6,-2.4) node {$J_{-}$};
\draw (2.6,2.4) node {$J_{+}$};
\draw (0,0) node {$H$};
\draw (3,0) node {$S$};
\draw [decoration={aspect=0, segment length=1.6mm, amplitude=0.7mm,coil},decorate] (-0.35,0)   -- (-1.5,0)node {\tiny$|$};
 \draw [decoration={aspect=0.4, segment length=1mm, amplitude=1mm,coil},decorate] (0,0.35) .. controls (0,1.2) and (1.2,1.5) .. (1.7,1.35);
  \draw [decoration={aspect=0.4, segment length=1mm, amplitude=1mm,coil},decorate] (0,-0.35) .. controls (0,-1.2) and (1.2,-1.5) .. (1.7,-1.35);
 \draw [decoration={aspect=0.4, segment length=1mm, amplitude=1mm,coil},decorate] (2.7,-1.2) -- (3,-0.53);
  \draw [decoration={aspect=0.4, segment length=1mm, amplitude=1mm,coil},decorate] (2.7,1.2) -- (3,0.53);
\end{tikzpicture}
\caption{General pinch surface corresponding to logarithmic divergences. Only gluons connecting the hard and soft to the jet surfaces can be present to give this divergence.}\label{pinch2}
\end{figure}

The next step towards the factorization of the $qq\gamma$ amplitude is to power-count (this amounts to just counting the power of the so called normal variables, i.e. the ones whose vanishing defines the pinch surface, in the numerator and in the denominator of the integrand) and find the most divergent solutions to Landau's equations. In \cite{Sterman1} it is proven that in any covariant gauge, the divergences are logarithmic at worst and the general pinch surface is of the form presented in Fig. \ref{pinch2} in $d=4$ dimensions. Where the jet functions $J_\pm$ are only connected to the hard function $H$ through one fermion line and longitudinally polarized gluons, the jets are connected to each other through soft gluons attached to the soft function $S$ \cite{Collins}. In all physical gauges the divergences are also logarithmic at worst and furthermore no lines except for the fermionic lines connect $H$ to the jets $J_\pm$ \cite{Stermannotes}.

Our goal now is to factorize the general pinch surface in Fig. \ref{pinch2} into contributions where in each of the regions ($H$, $J_\pm$ and $S$) the loop momenta are not restricted anymore in a gauge independent way (recall that in the soft and jet subdiagrams all the lines are soft and collinear respectively). This is performed through the introduction of the so called Wilson lines, which are related with the parallel transport in fibre bundles and the uniqueness of the so called ``horizontal lift'' \cite{Nakahara} of a spacetime curve $\gamma(t)$ with $t\in[t_1,t_2]$ between the two points $y=\gamma(t_1)$ and $z=\gamma(t_2)$
\begin{equation}
\Phi(t_2,t_1)\equiv \mathcal{P} \Big\{\exp\Big(-ig\mu^{\epsilon}\int_{t_1}^{t_2}dt\frac{d\gamma^\mu}{dt} A_\mu(\gamma(t))\Big)\Big\}\;,\label{Wilsonline}
\end{equation}
where the symbol $\mathcal{P}$ is the path ordering operator which orders the $A_\mu(\gamma(t))$ so that the ones with higher $t$ stand to the left (remember that the $A$'s are matrices). Basically, the Wilson lines help in building gauge-invariant objects between two different points in space-time.\par

The Wilson lines are introduced because they reproduce order by order the so called eikonal Feynman rule of soft gluon emissions. This is so because soft gluons only couple to the color and the direction of the jet they attach to. For this reason we will regard their interaction as eikonal.\par
 How to relate this fact to the Wilson line? Recall that internal lines in reduced diagrams are in free-flight so that their velocity $v^\mu=d\gamma^\mu(t)/dt$ is constant along their trajectory and we can hence write (\ref{Wilsonline}) for these particles as
\begin{equation}
\Phi_v(t_2,t_1)=\mathcal{P} \Big\{\exp\Big(-ig\mu^{\epsilon}\int_{t_1}^{t_2}dtv^\mu A_\mu(tv)\Big)\Big\}\;.
\end{equation}
The usual assumption in QFT is that the interaction (the qq$\gamma$ vertex in our case) occurs inside the collider experiment (set at the origin) and the final state particles travel out to infinity where they are detected. Hence we will treat with the special Wilson line $\Phi(\infty,0)$ and, expressing $A^\mu(x)$ as the sum of its Fourier coefficients, we see that it equals
\begin{equation}
\Phi_v(\infty,0)=\mathcal{P} \Big\{\exp\Big(-ig\mu^{\epsilon}\int_{0}^{\infty}dtv^\mu\int\frac{d^dk}{(2\pi)^d}\tilde{A}_\mu(k)e^{i t k\cdot v}\Big)\Big\} \;.
\end{equation}
To carry out the integration in $t$ we will Wick rotate $k^0\to ik^0$ so that the contribution from $t=\infty$ vanishes. In this way, after going back to real energy, we obtain
\begin{align}
\Phi_v(\infty,0)=\exp\Big(g\mu^{\epsilon}\int\frac{d^dk}{(2\pi)^d}\frac{v^\mu}{k\cdot v}\tilde{A}_\mu(k)\Big)\;,
\end{align}
which reproduces order by order the eikonal soft gluon emission Feynman rule from a fermion line (which is ${v^\mu}/{k\cdot v}$).\par
\begin{figure}[ht!]
\centering
\begin{tikzpicture}

\shadedraw[inner color=white,outer color=gray,draw=black,rotate=30]  (3.5,-0.) ellipse (20pt and 10pt);

\shadedraw[inner color=white,outer color=gray,draw=black,rotate=-30]  (3.5,-0.) ellipse (20pt and 10pt);
\shadedraw[inner color=white,outer color=gray,draw=black,rotate=-30]  (0,0) ellipse (10pt and 10pt);
\shadedraw[inner color=white,outer color=gray,draw=black,rotate=0]  (4.8,0) ellipse (10pt and 10pt);
\draw [thick](0.3,-0.2) -- (1.125,-0.675);
\draw [thick](0.3,0.2) -- (1.125,+0.675);

\draw [thick](1.6,-0.92) -- (4.05,-2.358);
\draw [thick](1.6,0.92) -- (4.05,2.358);
\draw [double](1.6,-0.88) -- (2.9,-0.2);
\draw [double](1.6,0.88) -- (2.9,0.2);
\draw (0,0) node {$H$};
\draw (4.8,0) node {$S$};
 \draw (1.5,-0.893) arc [start angle=180, end angle=-180, radius=0.05cm];
  \draw (1.5,0.893) arc [start angle=180, end angle=-180, radius=0.05cm];
\draw (2,-2) node {$J_-/\mathcal{J}_-$};
\draw (2,2) node {$J_+/\mathcal{J}_+$};
\draw [decoration={aspect=0, segment length=1.6mm, amplitude=0.7mm,coil},decorate] (-0.35,0)   -- (-1.5,0)node {\tiny$|$};

\draw [decoration={aspect=0.4, segment length=1mm, amplitude=1mm,coil},decorate] (4.1,0.3)   -- (4.47,0.1);
\draw [decoration={aspect=0.4, segment length=1mm, amplitude=1mm,coil},decorate] (4.1,-0.3)   -- (4.47,-0.1);
\draw [decoration={aspect=0.4, segment length=1mm, amplitude=1mm,coil},decorate] (4.75,-0.685)   -- (4.67,-0.325);
\draw [decoration={aspect=0.4, segment length=1mm, amplitude=1mm,coil},decorate] (4.75,0.685)   -- (4.67,0.325);

 \draw (3.6,0) arc [start angle=180, end angle=-180, radius=0.05cm];
 \draw [double](3.69,0.032) -- (5.05,0.9);
  \draw [double](3.69,-0.032) -- (5.05,-0.9);

\draw [decoration={aspect=0.4, segment length=1mm, amplitude=1mm,coil},decorate]  (3.2,-1.43) .. controls (3.2,-0.5) and (2.7,-0.6) .. (2.35,-0.5);
\draw [decoration={aspect=0.4, segment length=1mm, amplitude=1mm,coil},decorate]  (3.2,1.43) .. controls (3.2,0.5) and (2.7,0.6) .. (2.35,0.5);
\end{tikzpicture}
\caption{Factorized quark EM form factor corresponding to equation (\ref{factorization}). The double lines represent Wilson lines.}\label{factorized}
\end{figure}
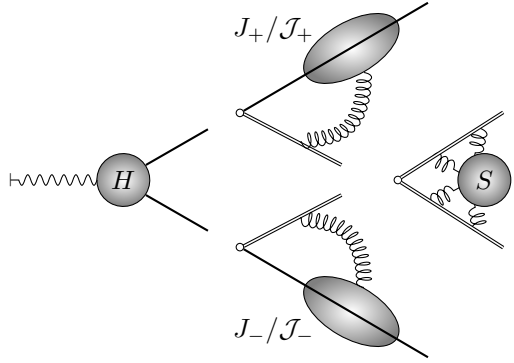

Thanks to the introduction of the Wilson lines, the so called eikonal identity (which represents the fact that at leading order in softness soft emissions factorize and are expressed in terms of independent emissions with eikonal vertices), and the use of Ward identities, it is possible to factorize the all order quark EM form factor \cite{Collins}. This will lead to express the general pinch surface in Fig. \ref{pinch2} in the factorized form in Fig. \ref{factorized}.\par
To get an idea how factorization will come about, notice that the interaction between two jets through soft gluons can be completely encoded in the soft function, defined as
\begin{equation}
\mathcal{S}(\beta_+\cdot \beta_-,\alpha_s(\mu^2),\epsilon)\equiv\bra{0} \Phi_{\beta_+}(\infty,0) \Phi_{\beta_-}(\infty,0)\ket{0}\;,\label{softf}
\end{equation}
where $\beta_+$ and $\beta_-$ are velocities proportional to the jet momenta $p_+$ and $p_-$ respectively, $\mu^2$ is a renormalization scale, $\epsilon$ the dimensional regulator, and $\alpha_s(\mu^2)$ is the renormalization-scale-dependent strong-force coupling-constant. \par
Now we want to describe how a fermion travels to the final state while interacting through soft gluons with the other fermion and hence define each jet leg as
\begin{equation}
J_\pm\Big(\frac{(p_\pm\cdot n_\pm)^2}{n_\pm^2\mu^2},\alpha_s(\mu^2),\epsilon\Big)u(p_\pm)={\bra{0} \Phi_{n_\pm}(\infty,0) \psi(0)\ket{p_\pm}}\,,\label{JET}
\end{equation}
where $\psi$ is the fermion field operator and $n_i$ is the direction of the Wilson line. To avoid spurious collinear singularities it is customary to choose $n_i^2\neq 0$ (although there are considerable computational advantages in setting $n^2=0$ with an easy fixing of the spurious singularities \cite{Bonocore:2015esa}).\par

Now we need to take into account the overlap of the soft and collinear regions to avoid double counting of divergences and also cancel graphs with eikonal self interactions. The overlapping can be seen as a jet function whose collinear gluons become soft or a soft function whose gluons become collinear. In either case, it can be described by the \textit{eikonal jet function} defined as
\begin{equation}
\mathcal{J}_i\Big(\frac{(\beta_i\cdot n_i)^2}{n_i^2\mu^2},\alpha_s(\mu^2),\epsilon\Big)\equiv{\bra{0} \Phi_{n_i}(\infty,0) \Phi_{\beta_i}(\infty,0)\ket{0}}\,.\label{eikonaljet}
\end{equation}
 Since we will divide the jet function (\ref{JET}) by (\ref{eikonaljet}), the eikonal self interactions of the Wilson lines $\Phi_{n_i}$ and $\Phi_{\beta_i}$ in (\ref{JET}) and (\ref{softf}) are canceled out. 
Defining the hard function $\mathcal{H}$ as the result of dividing the form factor $\Gamma$ by $\mathcal{S}\prod_i(J_i/\mathcal{J}_i)$, we can finally write the formula for the factorized form factor
\begin{align}
\Gamma\Big(\frac{\mu^2}{Q^2},\alpha_s(\mu^2),\epsilon\Big)&=\mathcal{H}	\Big(\frac{\mu^2}{Q^2},\alpha_s(\mu^2),\epsilon\Big)\mathcal{S}(\beta_+\cdot \beta_-,\alpha_s(\mu^2),\epsilon) \prod_{i=\pm}\frac{J_i\Big(\frac{(p_i\cdot n_i)^2}{n_i^2\mu^2},\alpha_s(\mu^2),\epsilon\Big)}{\mathcal{J}_i\Big(\frac{(\beta_i\cdot n_i)^2}{n_i^2\mu^2},\alpha_s(\mu^2),\epsilon\Big)}\;.\label{factorization}
\end{align}
Here it is implied that, since the soft and jet functions present in (\ref{factorization}) can generate spurious UV divergences, UV counterterms are introduced to cancel them.  Note that the functions defined to obtain (\ref{factorization}) depend only on general properties of the external particles like spin, charge or color, and collect all soft and collinear divergences. This dependence and some issues concerning the so called \textit{cusp anomaly} are studied in detail in \cite{DGM}. 
The factorization formula (\ref{factorization}) can be extended to more generic amplitudes. In cases with more legs, the color dependence of the amplitude is non-trivial but remains tractable. The presence of the so called Glauber gluons (see \cite{Collins} in pg. 14 for details) might spoil factorization. This is still an open topic of research (see for example \cite{CFR}-\cite{FSS} for more recent and refined results). Al things said, for fixed-angle scattering and the form-factor discussed here it is always possible to deform contours away from the Glauber region \cite{Collins}.

\section{Coordinate space factorization}\label{coordfact}
The momentum space results presented above were extensively studied from the early 1980's to the present day, however, the coordinate space analogue results appeared only recently thanks to O. Erdogan's work \cite{ErdoganCS}. There, Erdogan presents a precise derivation of the factorization formulas for the quark EM form factor at LP using the same steps as presented above. \par
 First, the unavoidable singularities of coordinate space Feynman graphs,
\begin{equation}
I(\{x_i^\mu\})=\prod_{\substack{\text{vertices}\\k}}\int d^dy_k\prod_{\substack{\text{lines}\\j}}\frac{F(\{x_i\},\{y_k\})}{[z_j^2+i\eta]^{p_j}}\;,\label{fefo}
\end{equation}
 are identified using again the Coleman-Norton interpretation (here the $\{y_k^\mu\}$ are the position of internal vertices and $\{x_i^\mu\}$ the positions of external points, and $z_j^\mu$ denotes the argument of the denominator in propagator of line $j$. $F(\{x_i\},\{y_k\})$ is a numerator factor containing all color factors,  constant and numerator factors (such as vertex derivative terms), and $p_j$ depends on whether the line is fermionic or bosonic).\par
 
For a general massless diagram, the pinched vanishing of the denominator of eq. (\ref{fefo}) defines Landau's equations in coordinate space
\begin{empheq}[left=\empheqlbrace]{align}
\sum_{\substack{\text{lines } j\\ \text{at vertex } k}}\alpha_jz_j^\mu\epsilon_{kj}=0 &\;\;\;\;\text{and} \nonumber\\
  \alpha_j\,z^2_j=0 & \;\;.
\end{empheq} 
It is of course intended that these equations must be satisfied together with the vanishing of the overall denominator obtained after Feynman parametrization. To connect with the Coleman-Norton picture, identify the product $\alpha_j z_j^\mu$ with a momentum vector $l^\mu\equiv \lambda\alpha_j z^\mu_j$ and $\lambda\alpha_j\equiv l_j^0/z^0_j$. Note that in the coordinate space picture the $\lambda$ parameter is interpreted in the inverse way as in the momentum space picture. This means that the soft singularities will have $\alpha_j=0$ or $\lambda$ going to zero. For a hard singularity, i.e. $z^0_j=|\vec{z}_j|\to0$, the Landau's Equations are satisfied automatically and $\lambda$ helps the $\alpha$ parameters remain finite. 
In this way we can see that each pinch singularity corresponds to massless particles propagating a finite distance on the lightcone between vertices with their momenta satisfying momentum conservation at each vertex. On the other hand off-lightcone particles have zero displacement when in divergent configurations \cite{ErdoganCS}. \par
If one studies the pinch surfaces for the $\gamma q q$ vertex one can easily recognize the same surfaces as in the momentum picture \cite{ErdoganCS}: two jets, one soft and one hard surface. Using power-counting techniques it is possible to see that the singularities of the vertex are,  at worst, logarithmic in $d=4$ dimensions and they correspond to the coordinate space analogue of Fig. \ref{pinch2}, where only gluons connect the hard and soft with jet surfaces respectively \cite{ErdoganCS}. \par
Finally, after using the so called hard-collinear and soft-collinear approximations it is also possible to see that the jets are connected to the soft and hard surfaces only by longitudinally polarized gluons (also known as scalar gluons) in Landau Gauge. Due to this fact, Ward identities are used to factorize the coordinate space vertex function as in Fig. \ref{factorized} \cite{ErdoganCS}.\par
These results help to factorize in the same fashion cross sections for partonic processes such as Drell-Yan (see \cite{Contopanagos:1996nh}). In \cite{Sterman:2016zby} it is demonstrated how, given Hermiticity of the interaction and using the Largest Time Equation \cite{Veltman2}, the cancellation of IR divergencies in inclusive cross sections comes to the fore in coordinate space. 
\section{Coordinate space explicit computations}\label{computations}
Next we will turn to compute some contributions to the jet function in coordinate space. To be able to compare our results we present here the results from momentum space calculations: \par
\subsection{Momentum space results for the one-loop jet function}
\begin{figure}[ht!]
\centering
\includegraphics[width=140mm]{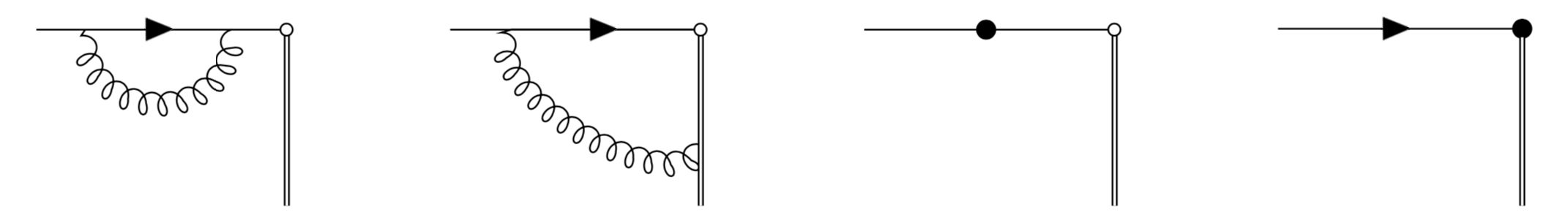}
\caption{Diagrams contributing to the first-order momentum space jet function. The solid dots represent UV counterterms and the double line is a Wilson line.}\label{jetff}
\end{figure}

For the jet function at one loop we will have the contributions from a self energy correction $J_p^{(1)}$ of the quark line and a gluon exchange vertex correction $J_V^{(1)}$ between the Wilson and the fermion line. All these contributions including the UV counterterms are presented in Fig. \ref{jetff}. The eikonal self interaction graphs do not appear due to the fact that we divided the jet function by the eikonal jet function in (\ref{eikonaljet}).

The first diagram in Fig. \ref{jetff} vanishes in dimensional regularization since it is scaleless  (remember the external momentum is lightlike, $p^2=0$) and there is no available quantity with non-zero mass dimension. This comes from a cancellation of UV and IR poles. Therefore, after introducing the UV counterterm in the third diagram of Fig. \ref{jetff}, we will obtain that $J_p^{(1)}=1/\epsilon_{\rm UV}$ (here I make explicit the UV nature of the pole in $\epsilon$). The vertex correction in the second diagram in Fig. \ref{jetff} amounts to
\begin{align}
J_V^{(1)}&=2 i\mu^{2\epsilon}g^2\int\frac{d^dk}{(2\pi)^d}\frac{(\slashed{k}-\slashed{p})\slashed{n}}{(k^2-i\eta)(k^2-2p\cdot k-i\eta)(2n\cdot k-i\eta)}\,,\label{eq:jetfdiagram}
\end{align}
where $n$ is the direction of the Wilson line and $n^2=0$. In the $\overline{\text{MS}}$ scheme, by using Dirac's equation on the spinor in the LHS of eq. (\ref{JET}),  adding the appropriate $\overline{\text{MS}}$ counterterm in the fourth diagram in Fig. \ref{jetff}  to eq. (\ref{eq:jetfdiagram}) 
 results in (see eq. (3.2) in \cite{Bonocore:2015esa}),
\begin{equation}
J_{V\rm r}^{(1)}\equiv J_V^{(1)}+J_{V,\rm CT}^{(1)}=2\left(\frac{\alpha_s}{4\pi }\right)\Big[\frac{1}{\epsilon^2}+\frac{1}{\epsilon}\Big(1-\gamma_E+\log\frac{4\pi\mu^2}{(-2p\cdot n)}\Big)+\mathcal{O}(\epsilon^0)\Big]\;.\label{eq:22}
\end{equation}

Now we turn to reproduce this result in coordinate space.

 \subsection{One-loop jet function in coordinate space}
 Once we have seen that factorization of the vertex function in coordinate space comes along pretty much as in the momentum space picture (see \cite{ErdoganCS}), let us now compute the one-loop jet function in coordinate space and see if, using the LSZ reduction formula, we can recover the results known in momentum space. \par
 
 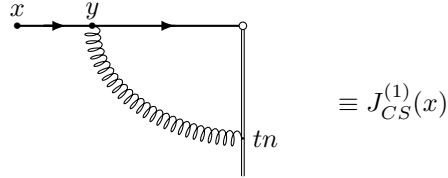
\begin{figure}[ht!]
 \centering
 \begin{tikzpicture}[>=stealth]
 \draw [decoration={aspect=0.4, segment length=1mm, amplitude=1mm,coil},decorate] (1,0) .. controls (1,-0.9) and (2.1,-1.5) .. (3,-1.5);
 \draw[-latex,thick] (2,0)--(2.1,0);
  \draw[thick] (2,0)--(2.95,0);
  \draw[-latex,thick] (0.35,0)--(.65,0);
   \draw[thick] (0.35,0)--(1,0);
    \draw [thick](0,0)--(0.55,0);
  \draw [thick](1,0)--(2.1,0);
   \draw (1,-0.00) node {\tiny$\bullet$};
      \draw (1,0.2) node {$y$};
          \draw (0,0.2) node {$x$};
          \draw (3.3,-1.5) node {$tn$};
          \draw (5,-1) node {$\equiv J^{(1)}_{CS}(x)$};
          
      \draw (0,-0.00) node {\tiny$\bullet$};
   \draw[double] (3,-0.05)--(3,-2);
   \draw (2.95,0) arc [start angle=180, end angle=-180, radius=0.05cm];
   \draw (3,-1.51) node {$\cdot$};
\end{tikzpicture}
\caption{Diagram representing the one-loop jet function (without UV conterterms) in coordinate space.}\label{repera}
\end{figure}

Reading from Fig. \ref{repera} we have that, taking the Wilson line to be lightlike ($n^2=0$),
\begin{align}
J^{(1)}_{CS}&(x)=(-ig T^a\gamma_\mu)\frac{\Gamma(1-\epsilon)\Gamma^2(2-\epsilon)}{(2\pi^{2-\epsilon})^24\pi^{2-\epsilon}}\int d^d y \frac{-(\slashed{x}-\slashed{y})}{((x-y)^2+i\eta)^{2-\epsilon}}\frac{-\slashed{y}}{(y^2+i\eta)^{2-\epsilon}}\int_0^{+\infty}dt \frac{(-igT^a)n^\mu}{((y-tn)^2+i\eta)^{1-\epsilon}}\,.\label{tonaser}
\end{align}
We can introduce Feynman parameters in two steps combining firstly the second and third denominators in the integrands of (\ref{tonaser}) and secondly the resulting denominator with the first one of (\ref{tonaser}) to identify the different configurations giving rise to unavoidable divergences. We will also employ this Feynman parametrization to solve the whole integral. In this case, using the delta functions in the Feynman parameters, Landau's equations read 
\begin{align}
&\alpha_2(x-y)^2+(1-\alpha_2)\alpha_1 y^2 +(1-\alpha_2)(1-\alpha_1)(y^2-2tn\cdot y)=0\;,\\
&\alpha_2(x-y)^\mu+(1-\alpha_2)\alpha_1 y^\mu+(1-\alpha_2)(1-\alpha_1)(y^\mu-tn^\mu)=0\,.
\end{align}
To study the case when the gluon line becomes soft we take $\alpha_1=1$. In this case we do not expect the gluon to change the direction of the external line since it carries no momentum. This is indeed the case because the solution to Landau's equations tells us that 
\begin{equation}
y^\mu=\frac{\alpha_2}{2\alpha_2-1} x^\mu\,,\label{EQ:222}
\end{equation}
and that $x$ and $y$ must lie in the lightcone for $\alpha_2\neq 0,1$.
For the end-point singularities in the $\alpha$ parameters we see that for $\alpha_2=1$ the vertex $y$ migrates to the external point $x$. This makes sense since the fermion line from $0$ to $y$ will be carrying all the momentum while the fermion line from $y$ to $x$ shrinks. Taking $\alpha_2=0$ means that the $y$ vertex migrates to the origin, signaling that we are dealing with an UV divergence. However in both cases there is not restriction to $y$ or $x$ to lie on the lightcone.
\par
For the case when $\alpha_1=0$, so that the exchanged gluon is not soft, Landau's equations tell us that
\begin{equation}
y^\mu=\frac{(1-\alpha_2)tn^\mu-\alpha_2 x^\mu}{(1-2\alpha_2)}\;.
\end{equation}
As we will see below, the only contribution to this amplitude will come when $t=0$ so that we will have a collinear pinch surface with the vertex $y$ obeying (\ref{EQ:222}) but this time the gluon emerges from the Wilson line cusp (the origin), making the gluon collinear. Both the fermion external line and gluon line are lightlike for $\alpha_2\neq0,1$. Again if $\alpha_2=1$ the internal vertex $y$ coincides with the external point $x$ and if $\alpha_2=0$ we will have that $y^\mu=tn^\mu$ signaling again a UV divergence for $t=0$.
\par
To carry out the whole integral we will introduce Feynman parameters in two steps as above, first combining the second and third denominators and then the resulting denominator with the first one in (\ref{tonaser}). In this way
\begin{align}
J^{(1)}_{CS}(x)&=-g^2 C_F\frac{\Gamma(1-\epsilon)\Gamma^2(2-\epsilon)}{16\pi^{6-3\epsilon}}\frac{\Gamma(5-3\epsilon)}{\Gamma^2(2-\epsilon)\Gamma(1-\epsilon)}\times\nonumber\\
&\times\int d^d y\int_0^{+\infty}dt\int_0^1 d\alpha_1d\alpha_2\frac{(\slashed{x}\slashed{n}\slashed{y}-\slashed{y}\slashed{n}\slashed{y})\alpha_1^{1-\epsilon}(1-\alpha_1)^{-\epsilon}\alpha_2^{1-\epsilon}(1-\alpha_2)^{2-2\epsilon}}{(y^2-2y\cdot(\alpha_2 x+(1-\alpha_1)(1-\alpha_2)tn)+\alpha_2 x^2+i\eta)^{5-3\epsilon}}\;.
\end{align}
Now we shift the integration variable $y\to y-(\alpha_2 x+(1-\alpha_1)(1-\alpha_2)tn)$ and, by parity, drop odd powers of the new $y$ in the numerator. This yields
\begin{align}
&J^{(1)}_{CS}(x)=-g^2 C_F\frac{\Gamma(5-3\epsilon)}{16\pi^{6-3\epsilon}}\int d^d y\int_0^{+\infty}dt\int_0^1 d\alpha_1d\alpha_2\frac{(\alpha_2(1-\alpha_2)\slashed{x}\slashed{n}\slashed{x}-\slashed{y}\slashed{n}\slashed{y})\alpha_1^{1-\epsilon}(1-\alpha_1)^{-\epsilon}\alpha_2^{1-\epsilon}(1-\alpha_2)^{2-2\epsilon}}{(y^2+\alpha_2(1-\alpha_2) x^2-2(1-\alpha_1)(\alpha_2-\alpha_2^2)tn\cdot x+i\eta)^{5-3\epsilon}}\;.
\end{align}
At this point, it is possible (taking $n\cdot x\neq0$) to carry out the integration in $t$ (giving only a contribution at $t=0$). This gives
\begin{align}
&J^{(1)}_{CS}(x)=g^2 C_F\frac{\Gamma(5-3\epsilon)}{16\pi^{6-3\epsilon}}\int d^d y\int_0^1 d\alpha_1d\alpha_2\frac{(\alpha_2(1-\alpha_2)\slashed{x}\slashed{n}\slashed{x}-\slashed{y}\slashed{n}\slashed{y})\alpha_1^{1-\epsilon}(1-\alpha_1)^{-1-\epsilon}\alpha_2^{-\epsilon}(1-\alpha_2)^{1-2\epsilon}}{(4-3\epsilon) 2n\cdot x(y^2+\alpha_2(1-\alpha_2) x^2+i\eta)^{4-3\epsilon}}\;.
\end{align}
Using standard Dimensional Regularization formulas we carry out the integral in $y$ and identify the Euler Beta functions in the Feynman parameters integrations to get
\begin{align}
J^{(1)}_{CS}&(x)=\frac{-i\pi^{\frac{d}{2}}g^2 C_F(x^2)^{2\epsilon-2}}{(2x\cdot n) 16\pi^{6-3\epsilon}}\frac{\Gamma(1-2\epsilon)\Gamma(2-\epsilon)\Gamma(-\epsilon)}{\Gamma(2-2\epsilon)}\Big(\frac{\Gamma(\epsilon)(1+\epsilon)}{\Gamma(2+\epsilon)}\Big)\Big(2x\cdot n \slashed{x}(1-2\epsilon)+\epsilon\slashed n x^2\Big)\;.\label{JCs}
\end{align}

The consequence of expanding in powers of $\epsilon$ the result in eq. (\ref{JCs}) is,
\begin{align}
J^{(1)}_{CS}(x)=&\frac{i \alpha_s C_F}{4\pi}\frac{-\slashed x}{2\pi^2(x^2+i\eta)^2}\Big(-\frac{2}{\epsilon^2}+\frac{2(1-2\gamma_E)}{\epsilon}-(\slashed x)^{-1}\frac{\slashed n x^2}{n\cdot x}\frac{1}{\epsilon}\Big)+\mathcal{O}(\epsilon^0)\;.\label{eq:expandedresult}
\end{align}

This is a very nice result since the most divergent part is proportional to a Fermion propagator $S(x)$ (one can think of this result as follows: the gluon merges collinearly with the fermion producing the divergence times the fermion propagator) and this will allow us to use the LSZ formula to get the momentum space expression in a very simple manner. This formula relates the jet function in momentum space, $J_{MS}$, with the one in coordinate space as
\begin{align}
J_{MS}(p)=-i\int {d^dx} e^{-ip\cdot x} (i\slashed\partial)J^{(1)}_{CS}(x)\;.
\end{align}
By using the Dirac equation $\slashed\partial S(x)=i\delta^{(d)}(x)$ we find that the most divergent part of the jet function is
\begin{align}
\lim_{\epsilon\to0}J_{MS}(p)=\lim_{\epsilon\to0}\frac{\alpha_s C_F}{4\pi} \Big(\frac{2}{\epsilon^2}\Big)\;.
\end{align}
So that we recover the same leading pole structure as the one-loop gluon exchange vertex correction $J_{V\rm r}^{(1)}$ of eq. (\ref{eq:22}). The last subleading term inside the brackets (\ref{eq:expandedresult}) must contain the dependence on the collinear scale $p\cdot n$ after LSZ reduction (which is not straightforward to perform). \par

\subsection{Abelian radiative one-loop jet function}\label{radiativesection}

\begin{figure}[ht!]
 \centering
 \begin{tikzpicture}[>=stealth]
 \draw [decoration={aspect=0.4, segment length=1mm, amplitude=1mm,coil},decorate] (1,0) .. controls (1,-0.9) and (2.1,-1.5) .. (3,-1.5);
  \draw [decoration={aspect=0.4, segment length=1mm, amplitude=1mm,coil},decorate] (.2,1) -- (2,0);
 \draw[-latex,thick] (2.3,0)--(2.6,0);
 \draw[-latex,thick] (1.25,0)--(1.5,0);
  \draw[-latex,thick] (0.35,0)--(.6,0);
    \draw [thick](0,0)--(2.95,0);
  \draw [thick](1,0)--(2.1,0);
   \draw (1,-0.00) node {\tiny$\bullet$};
   \draw (1.98,-0.00) node {\tiny$\bullet$};
      \draw (.2,0.99) node {\tiny$\bullet$};
      \draw (0,-0.00) node {\tiny$\bullet$};
      \draw (1.5,-2.00) node {(a)};
   \draw[double] (3,-0.05)--(3,-2);
   \draw (2.95,0) arc [start angle=180, end angle=-180, radius=0.05cm];
   \draw (3,-1.51) node {$\cdot$};
\end{tikzpicture}\;\;\;\;
 \begin{tikzpicture}[>=stealth]
 \draw [decoration={aspect=0.4, segment length=1mm, amplitude=1mm,coil},decorate] (2,0) .. controls (2,-0.9) and (2.8,-.9) .. (3,-.9);
 \draw (1.5,-2.00) node {(b)};
  \draw [decoration={aspect=0.4, segment length=1mm, amplitude=1mm,coil},decorate] (.2,1) -- (1,0);
 \draw[-latex,thick] (2.3,0)--(2.6,0);
 \draw[-latex,thick] (1.25,0)--(1.65,0);
  \draw[-latex,thick] (0.35,0)--(.6,0);
    \draw [thick](0,0)--(2.95,0);
  \draw [thick](1,0)--(2.1,0);
   \draw (1,-0.00) node {\tiny$\bullet$};
   \draw (1.98,-0.00) node {\tiny$\bullet$};
      \draw (.2,0.99) node {\tiny$\bullet$};
      \draw (0,-0.00) node {\tiny$\bullet$};
   \draw[double] (3,-0.05)--(3,-2);
   \draw (2.95,0) arc [start angle=180, end angle=-180, radius=0.05cm];
   \draw (3,-.91) node {$\cdot$};
\end{tikzpicture}\;\;\;\;
\begin{tikzpicture}[>=stealth]
\draw (1.5,-2.00) node {(c)};
 \draw [decoration={aspect=0.4, segment length=1mm, amplitude=1mm,coil},decorate] (0.75,0) .. controls (0.75,-0.9) and (2.25,-.9) .. (2.25,0);
  \draw [decoration={aspect=0.4, segment length=1mm, amplitude=1mm,coil},decorate] (.2,1) -- (1.5,0);
  \draw[-latex,thick] (2.6,0)--(2.75,0);
  \draw[-latex,thick] (1,0)--(1.25,0);
    \draw[-latex,thick] (1.75,0)--(2,0);
  \draw[-latex,thick] (0.35,0)--(.6,0);
    \draw [thick](0,0)--(2.95,0);
  \draw [thick](1,0)--(2.1,0);
   \draw (0.75,-0.00) node {\tiny$\bullet$};
   \draw (2.25,-0.00) node {\tiny$\bullet$};
   \draw (1.5,-0.00) node {\tiny$\bullet$};
      \draw (.2,0.99) node {\tiny$\bullet$};
      \draw (0,-0.00) node {\tiny$\bullet$};
   \draw[double] (3,-0.05)--(3,-2);
   \draw (2.95,0) arc [start angle=180, end angle=-180, radius=0.05cm];
\end{tikzpicture}\;\;\;\;
 \begin{tikzpicture}[>=stealth]
 \draw (1.5,-2.00) node {(d)};
 \draw [decoration={aspect=0.4, segment length=1mm, amplitude=1mm,coil},decorate] (1.5,0) .. controls (1.5,-0.8) and (2.5,-.8) .. (2.5,0);
  \draw [decoration={aspect=0.4, segment length=1mm, amplitude=1mm,coil},decorate] (.2,1) -- (1,0);
  \draw[-latex,thick] (2.,0)--(2.15,0);
    \draw[-latex,thick] (2.5,0)--(2.9,0);
 \draw[-latex,thick] (1.,0)--(1.375,0);
  \draw[-latex,thick] (0.35,0)--(.6,0);
    \draw [thick](0,0)--(2.95,0);
  \draw [thick](1,0)--(2.1,0);
   \draw (1,-0.00) node {\tiny$\bullet$};
   \draw (1.5,-0.00) node {\tiny$\bullet$};
    \draw (2.5,-0.00) node {\tiny$\bullet$};
      \draw (.2,0.99) node {\tiny$\bullet$};
      \draw (0,-0.00) node {\tiny$\bullet$};
   \draw[double] (3,-0.05)--(3,-2);
   \draw (2.95,0) arc [start angle=180, end angle=-180, radius=0.05cm];
\end{tikzpicture}
\caption{Diagrams contributing to the one-loop radiative jet function in coordinate space. No external leg corrections are included.}\label{radiatedjetabelian}
\end{figure}
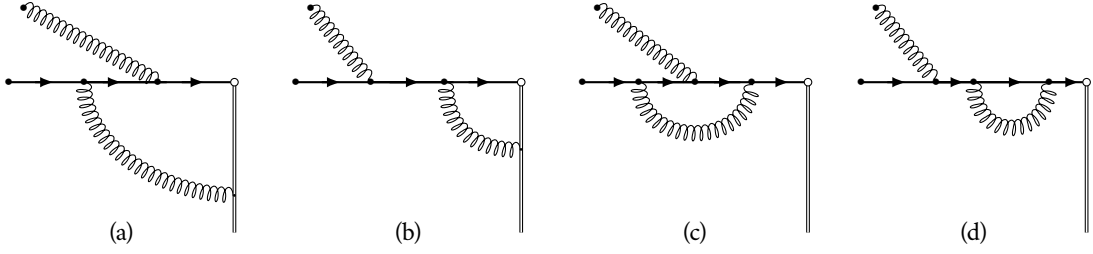

As stated in the introduction, the study of subleading regions (also called NLP regions, i.e. kinematical configurations that give rise to divergent subleading terms in observables, usually coming from emission of soft gluons) has gained attention in recent years \cite{Gervais:2017yxv,Beneke:2017ztn,Moult:2019mog,Laenen:2020nrt,Liu:2021mac}. For this reason, we will now turn on to analyze some features of the radiative one-loop jet function in coordinate space and try to reduce in quadrature the contributing diagrams. 
The contributions to the one-loop radiative jet function in the abelian case are listed in Figure \ref{radiatedjetabelian}. Here I do not include external leg corrections. 

In the interest regarding the study of NLP regions, we will also analyze Landau's equations for each diagram contributing to the radiative jet function. This will help us again to identify the configurations giving rise to collinear divergences and their overlapping with soft divergences of the emitted gluon (highlighted in the text below). To see this, as an oversimplified example, imagine that, after reducing to Feynman parameters a contribution to the radiative jet, one identifies through Landau's equations that $\alpha_1=0$ entails a collinear divergence and $\alpha_2=1$ a soft one. By separating the divergent part from the finite one on each of the two Feynman parameters (through the usual multiplication by $1=\alpha_i+(1-\alpha_i)$), one would get four terms
$\alpha_1\alpha_2+\alpha_1(1-\alpha _2) +\alpha_2 (1-\alpha_1) +(1-\alpha_1)(1-\alpha_2)$ encoding a collinear-finite, a collinear-soft overlapping, a finite-finite, and a finite-soft region respectively. This example shows the potentiality of the present treatment concerning the factorization and overlapping of divergent kinematical regions.

One should however be careful when analyzing divergences on a diagram by diagram basis, since results may suffer from artefacts due to the specific integral representation under treatment. All things said, the direct identification of individual Feynman diagrams with certain physical processes is a well established practice in theoretical particle physics. On the other hand, if one has access to the fully integrated final result in coordinate space, the soft and collinear limits can be studied by taking these in the values of the external momenta for the outgoing fermion and gluon (after performing LSZ reduction of the coordinate space, UV finite,  result).

\subsubsection{One-loop jet function with internal gluon emission in coordinate space}
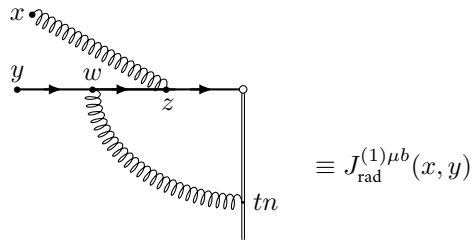
\begin{figure}[ht!]
 \centering
 \begin{tikzpicture}[>=stealth]
 \draw [decoration={aspect=0.4, segment length=1mm, amplitude=1mm,coil},decorate] (1,0) .. controls (1,-0.9) and (2.1,-1.5) .. (3,-1.5);
  \draw [decoration={aspect=0.4, segment length=1mm, amplitude=1mm,coil},decorate] (.2,1) -- (2,0);
 \draw[-latex,thick] (2.3,0)--(2.6,0);
 \draw[-latex,thick] (1.25,0)--(1.5,0);
  \draw[-latex,thick] (0.35,0)--(.6,0);
    \draw [thick](0,0)--(2.95,0);
  \draw [thick](1,0)--(2.1,0);
   \draw (1,-0.00) node {\tiny$\bullet$};
   \draw (1.98,-0.00) node {\tiny$\bullet$};
      \draw (.2,0.99) node {\tiny$\bullet$};
      \draw (1,0.2) node {$w$};
      \draw (2,-0.2) node {$z$};
      \draw (0,1) node {$x$};
          \draw (0,0.2) node {$y$};
          \draw (3.3,-1.5) node {$tn$};
          \draw (5,-1) node {$\equiv J^{(1)\mu b}_{\text{rad}}(x,y)$};
          
      \draw (0,-0.00) node {\tiny$\bullet$};
   \draw[double] (3,-0.05)--(3,-2);
   \draw (2.95,0) arc [start angle=180, end angle=-180, radius=0.05cm];
   \draw (3,-1.51) node {$\cdot$};
\end{tikzpicture}
\caption{Diagram representing the one-loop internal gluon emission from a jet.}\label{repera2}
\end{figure}

The first contribution to the abelian one-loop radiative jet function in Fig. \ref{radiatedjetabelian} is depicted in detail in Fig. \ref{repera2}. We consider a lightlike Wilson line, i.e. $n^2=0$. I will keep the non-abelian numerical factors (such as the fundamental Casimir of the gauge group, $C_F$).

This diagram amounts to
\begin{align}
J^{(1)\mu b}_{\text{rad}}(x,y)&=(ig^3T^aT^bT^a)\Big(\frac{\Gamma(2-\epsilon)}{2\pi^{2-\epsilon}}\Big)^3\Big(\frac{\Gamma(1-\epsilon)}{4\pi^{2-\epsilon}}\Big)^2
\int d^dzd^dw \int_0^\infty d t \frac{-(\slashed y-\slashed w)}{((w-y)^2+i\eta)^{2-\epsilon}} \frac{-\slashed n(\slashed w-\slashed z)}{((w-z)^2+i\eta)^{2-\epsilon}}\times\nonumber\\
&\times \gamma^\mu\frac{-\slashed z}{(z^2+i\eta)^{2-\epsilon}}\frac{1}{((w-tn)^2+i\eta)^{1-\epsilon}}\frac{1}{((x-z)^2+i\eta)^{1-\epsilon}}\;.\label{radjet1}
\end{align}
The color structure above is easily computable, it gives $T^aT^bT^a=-T^b/(2N)$ for $SU(N)$. This factor is also easily  computed for $SO(N)$, where is just $T^b/4$ (giving no suppression of the diagram for large number of colors), and for $Sp(N)$ where it amounts to $-T^b/4$ (see the Appendix in \cite{Llanes-Estrada:2018azk} for a detailed derivation).\par

\paragraph{Analysis of Landau's equations}
We will analyze Landau's equations for this diagram by anticipating the way we will use Feynman parameters to solve eq. (\ref{radjet1}). For the integration in $z$ we will first combine the second and third denominators and subsequently the resulting denominator with the last denominator in eq. (\ref{radjet1}). Landau's equations for the resulting denominator read
\begin{align}
&z^2-2z\cdot (\alpha_1\alpha_2 w+(1-\alpha_2)x)+\alpha_1\alpha_2 w^2+(1-\alpha_2)x^2=0\label{land1}\;,\\
&z^\mu=\alpha_1\alpha_2 w^\mu+(1-\alpha_2)x^\mu\label{land2}\;.
\end{align}
Let us analyze the case when $\boxed{\alpha_2=1}$, which corresponds to the case of a \textbf{soft emitted gluon} (since the Feynman parameter of the external gluon line is $1-\alpha_2$). In this case eq. (\ref{land2}) amounts to $z^\mu=\alpha_1 w^\mu$ so that the emitted gluon indeed does not change the direction of the fermion. Using this result in eq. (\ref{land1}) for a lightlike external point, $x^2=0$, one obtains $\alpha_1(1-\alpha_1)w^2=0$ which sets $w$ on the lightcone except for the endpoints of the integration in $\alpha_1$, these endpoints correspond to UV divergences (in two distinct fermion lines) $z^\mu=0$ for $\alpha_1=0$ and $z^\mu=w^\mu$ for $\alpha_1=1$. Next, let us take $\boxed{\alpha_2=0}$, this gives the UV condition that $x^\mu=z^\mu$ so that the emitted gluon travels no distance. 

Now, setting $\boxed{\alpha_1=1}$, eq. (\ref{land2}) tells us that $z$ must lie in the line connecting $x$ and $w$ since it equals $z^\mu=\alpha_2w^\mu+(1-\alpha_2)x^\mu$. Using this in eq. (\ref{land1}) one obtains the condition $ \alpha_2(1-\alpha_2)(w-x)^2=0$ which sets $w$ on the lightcone of $x$ except for the UV gluon with $\alpha_2=0$ and hence $z^\mu=x^\mu$ or the UV fermion with $\alpha_2=1$ and $z^\mu=w^\mu$. For the case $\boxed{\alpha_1=0}$ one has that $z^\mu=(1-\alpha_2)x^\mu$ which gives $\alpha_2(1-\alpha_2)x^2=0$, which is automatically fulfilled if the external point is taken on the lightcone.\\

Landau's equations for the other vertex $w$ in eq. (\ref{radjet1}), after the steps taken to reach eq. (\ref{intermediatestep}) below, read
\begin{align}
\alpha w^2+2w\cdot \theta+\kappa^2&=0\\
\alpha w^\mu+\theta^\mu&=0\;.
\end{align}
Here $\alpha=\alpha_3\alpha_4\alpha_5+\alpha_5(1-\alpha_3)\alpha_1\alpha_2(1-\alpha_1\alpha_2)+\alpha_5(1-\alpha_4)$, $ \theta^\mu=-(1-\alpha_5)n^\mu-\alpha_1\alpha_2(1-\alpha_2)x^\mu-\alpha_5(1-\alpha_4)y^\mu$, and $\kappa^2=\alpha_5(1-\alpha_3)\alpha_2(1-\alpha_2)x^2+\alpha_5(1-\alpha_4)y^2$. \\
We will set the external vertices to be on the lightcone $x^2=y^2=0$, which means $\kappa^2=0$. For the case $\boxed{\alpha\neq0}$ we will obtain that $\theta^2=0$ which gives the condition on the external vertices $(1-\alpha_5)n\cdot(-\alpha_1\alpha_2(1-\alpha_2)x-\alpha_5(1-\alpha_4)y)+\alpha_1\alpha_2(1-\alpha_2)\alpha_5(1-\alpha_4)x\cdot y=0$. Which can be solved for a collinear gluon emission with $x\propto y$ and $\alpha_5=1$. Another solution has the quarks and the gluon emitted from the vertex all traveling in the same direction, \textit{i.e.} $x,y\propto n$. The case $\boxed{\alpha=0}$ as seen from the result in eq. (\ref{eq:23}) entails several UV divergences arising from the fact that there is a Jacobian of $\alpha^{-2+\epsilon}$ (UV divergences are regulated for $\epsilon>0$). Some of them have $\alpha_4=\alpha_5=1$, $\alpha_3=0$ and $\alpha_1=\alpha_2=1$, $\alpha_1=\alpha_2=0$ or $\alpha_1=0$ and $\alpha_2=1$.

\paragraph{Computation} As in the previous subsection we introduce Feynman parameters in two steps for combining the second, the third, and last denominators above such that
\begin{align}
&J^{(1)\mu b}_{\text{rad}}(x,y)=F^b\int d^dw\int d^dz\int_0^\infty dt \int_0^1d\alpha_1d\alpha_2\frac{(\alpha_1^{1-\epsilon}(1-\alpha_1)^{1-\epsilon}\alpha_2^{3-2\epsilon}(1-\alpha_2)^{-\epsilon})}{((w-y)^2+i\eta)^{2-\epsilon}((w-tn)^2+i\eta)^{1-\epsilon}}\times\nonumber\\
&\times\frac{(\slashed y-\slashed w)\slashed n( \slashed z \gamma^\mu\slashed z-\slashed w\gamma^\mu \slashed z)}{(z^2-2z\cdot (\alpha_1\alpha_2 w+(1-\alpha_2)x)+\alpha_1\alpha_2 w^2+(1-\alpha_2)x^2)^{5-3\epsilon}}\;,
\end{align}
where $F^b$ encodes the color structure and numerical factors. Now, to compute the integral in $z$, we shift it as $z\to z-(\alpha_1\alpha_2 w+(1-\alpha_2)x)$ and then drop odd powers of $z$ in the numerator due to parity. This integration results in
\begin{align}
J^{(1)\mu b}_{\text{rad}}(x,y)&=F'^b\int d^dw\int_0^\infty dt \int_0^1d\alpha_1d\alpha_2\frac{\alpha_1^{1-\epsilon}(1-\alpha_1)^{1-\epsilon}\alpha_2^{3-2\epsilon}(1-\alpha_2)^{-\epsilon}}{((w-y)^2+i\eta)^{2-\epsilon}((w-tn)^2+i\eta)^{1-\epsilon}} \Big(K^\mu-(\slashed y-\slashed w)\slashed n \gamma^\mu (m^2)/2\Big)(m^2)^{2\epsilon-3}\;,
\end{align}
where $m^2=\alpha_1\alpha_2(1-\alpha_1\alpha_2)w^2+\alpha_2(1-\alpha_2)x^2-2\alpha_1\alpha_2(1-\alpha_2)  x\cdot w$ and $K^\mu=(\slashed y-\slashed w) \slashed n [((1+\alpha_1\alpha_2) \slashed w+(1-\alpha_2)\slashed x)\gamma^\mu(\alpha_1\alpha_2 \slashed w+(1-\alpha_2)\slashed x)$. Next we combine the second denominator above with the last one again (the one proportional to $m^2$) with Feynman parameters and then compute the integral in $t$ assuming we are not integrating $w$ in the time-like region where $w\cdot n\geq0$. From this integration we will obtain a denominator factor and a $-2n\cdot w$. We will combine these with the last propagator denominator left in two steps of Feynman parametrization, which gives
\begin{align}
J^{(1)\mu b}_{\text{rad}}(x,y)&=F''^b\int d^dw \int_0^1d\alpha_1...d\alpha_5{\alpha_1^{1-\epsilon}(1-\alpha_1)^{1-\epsilon}\alpha_2^{3-2\epsilon}(1-\alpha_2)^{-\epsilon}\alpha_3^{-1-\epsilon}(1-\alpha_3)^{2-2\epsilon}}\times\nonumber\\
&\times \frac{\alpha_4^{2-3\epsilon}(1-\alpha_4)^{1-\epsilon}\alpha_5^{4-4\epsilon}}{(\alpha w^2+2w\cdot  \theta+\kappa^2)^{6-4\epsilon}}\Big(K^\mu-(\slashed y-\slashed w)\slashed n \gamma^\mu (m^2)/2\Big)\;.\label{intermediatestep}
\end{align}
 Finally we shift the integration variable as $w\to w\alpha^{\frac{1}{2}}+ \theta\alpha^{-\frac{1}{2}}$ and carry out the $w$ integration finding
\begin{align}
J^{(1)\mu b}_{\text{rad}}(x,y)&=\frac{ig^3\pi^{3\epsilon-6}T^b}{2^8N(3-3\epsilon)}\times\int_0^1d\alpha_1...d\alpha_5\alpha_1^{1-\epsilon}(1-\alpha_1)^{1-\epsilon}\alpha_2^{3-2\epsilon}(1-\alpha_2)^{-\epsilon}\alpha_3^{-1-\epsilon}(1-\alpha_3)^{2-2\epsilon} \alpha_4^{2-3\epsilon}(1-\alpha_4)^{1-\epsilon}\times\nonumber\\
&\times\alpha_5^{4-4\epsilon}\alpha^{-2+\epsilon}\Big(\alpha g^{\rho\sigma}A^\mu_{\rho\sigma}\Gamma(3-3\epsilon)(M^2)/2+B^\mu\Gamma(4-3\epsilon)\Big)(M^2)^{3\epsilon-4}\;,\label{eq:23}
\end{align}
for $SU(N)$.
The expressions for the $A_{\rho\sigma}^\mu$, $B^\mu$, and $M^2$ are shown next (here $\beta^\mu\equiv\alpha^{-\frac{1}{2}}\theta^\mu$):
\begin{align}
M^2=\kappa^2-\beta^2\;,\label{M2}
\end{align}
\begin{align}
B^\mu=&(\slashed y-\slashed\beta)\slashed n\Big[ \big( (1+\alpha_1\alpha_2)\slashed\beta+(1-\alpha_2)\slashed x \big)\gamma^\mu \big( \alpha_1\alpha_2\slashed\beta +(1-\alpha_2)\slashed x\big)-\nonumber\\
&-\gamma^\mu \big( \alpha_1\alpha_2(1-\alpha_1\alpha_2)\beta^2+\alpha_2(1-\alpha_2)x^2-2\alpha_1\alpha_2(1-\alpha_2)\beta\cdot x\big)/2\Big]\,,
\end{align}
\begin{align}
A_{\rho\sigma}^\mu=&(\slashed y-\slashed\beta)\big((1+\alpha_1\alpha_2)\alpha_1\alpha_2\gamma_\rho\gamma^\mu\gamma_\sigma-\alpha_1\alpha_2(1-\alpha_1\alpha_2)\gamma^\mu g_{\rho\sigma}/2\big)-\nonumber\\
&-\gamma_\rho \slashed n\Big[ (1+\alpha_1\alpha_2)\gamma_\sigma\gamma^\mu\big((1-\alpha_2)\slashed x +\alpha_1\alpha_2 \slashed \beta \big)+\alpha_1\alpha_2\big((1+\alpha_1\alpha_2)\slashed \beta+(1-\alpha_2)\slashed x\big)\gamma^\mu\gamma_\sigma-\nonumber\\
&-\alpha_1\alpha_2\gamma^\mu\big((1-\alpha_1\alpha_2)\beta_\sigma-(1-\alpha_2) x_\sigma\big)\Big]\;.
\end{align}

This result is rather cumbersome and will be left for a future work to  fully solve the integrations over Feynman parameters in eq. (\ref{eq:23}). Nonetheless, by inspection, one straightforward possible divergence is identified by noticing that in eq. (\ref{eq:23}) there is a $\alpha_3^{-1-\epsilon}$ factor that will produce a divergence whenever $\alpha_3=0$. Since $\alpha_3$ multiplies the denominator $((w-tn)^2+i\eta)^{1-\epsilon}$, this limit can be identified with the configuration where the internal gluon line goes soft (indeed it is regulated for $\epsilon<0$). This is always the case for gluon lines, due to the power of their propagator. This happens too for the external gluon, which produces a soft divergence whenever $\alpha_2=1$ and overlaps with a possible collinear singularity whenever $\alpha_5=1$ at the same time.  \\
As already stated above, the Jacobian $\alpha^{-2+\epsilon}$ in eq. (\ref{eq:23}) encodes several UV divergences (regulated for $\epsilon>0$). Also, the appearance of the $M^2$ term in eq. (\ref{eq:23}) also helps identifying further divergent configurations: for example, $M^2$ vanishes for lightlike and collinear external points, $x^2=y^2=x\cdot y=0$, (in this case one needs to set also $\alpha_5=1$). 

\subsubsection{One-loop jet function with external gluon emission in coordinate space}
We will continue by reducing in quadrature the second contribution to the abelian one-loop radiative jet function in Fig. \ref{radiatedjetabelian}.
\begin{figure}[ht!]
 \centering
 \begin{tikzpicture}[>=stealth]
 \draw [decoration={aspect=0.4, segment length=1mm, amplitude=1mm,coil},decorate] (2,0) .. controls (2,-0.9) and (2.8,-.9) .. (3,-.9);
  \draw [decoration={aspect=0.4, segment length=1mm, amplitude=1mm,coil},decorate] (.2,1) -- (1,0);
 \draw[-latex,thick] (2.3,0)--(2.6,0);
 \draw[-latex,thick] (1.25,0)--(1.65,0);
  \draw[-latex,thick] (0.35,0)--(.6,0);
    \draw [thick](0,0)--(2.95,0);
  \draw [thick](1,0)--(2.1,0);
   \draw (1,-0.00) node {\tiny$\bullet$};
   \draw (1.98,-0.00) node {\tiny$\bullet$};
      \draw (.2,0.99) node {\tiny$\bullet$};
      \draw (0,-0.00) node {\tiny$\bullet$};
   \draw[double] (3,-0.05)--(3,-2);
   \draw (2.95,0) arc [start angle=180, end angle=-180, radius=0.05cm];
   \draw (3,-.91) node {$\cdot$}; 
   \draw (5,-1) node {$\equiv J^{(1)\mu b}_{\text{ext, rad}}(x,y)$};
         \draw (1,-0.2) node {$z$};
      \draw (2,0.2) node {$w$};
      \draw (0,1) node {$x$};
          \draw (0,0.2) node {$y$};
          \draw (3.3,-0.9) node {$tn$};
\end{tikzpicture}
\caption{Diagram representing the one-loop external gluon emission from a jet.}\label{repera33}
\end{figure}
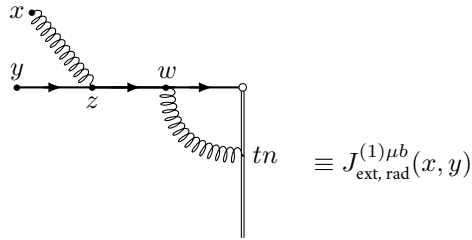

The diagram in Fig. \ref{repera33} equals
\begin{align}
J^{(1)\mu b}_{\text{ext, rad}}(x,y)&=(ig^3T^bC_F)\Big(\frac{\Gamma(2-\epsilon)}{2\pi^{2-\epsilon}}\Big)^3\Big(\frac{\Gamma(1-\epsilon)}{4\pi^{2-\epsilon}}\Big)^2
\int d^dzd^dw \int_0^\infty d t \frac{-(\slashed y-\slashed z)}{((z-y)^2+i\eta)^{2-\epsilon}} \frac{-\gamma^\mu(\slashed z-\slashed w)}{((z-w)^2+i\eta)^{2-\epsilon}}\times\nonumber\\
&\times \slashed n \frac{-\slashed w}{(w^2+i\eta)^{2-\epsilon}}\frac{1}{((w-tn)^2+i\eta)^{1-\epsilon}}\frac{1}{((x-z)^2+i\eta)^{1-\epsilon}}\;.\label{radjet2}
\end{align}
\paragraph{Analysis of Landau's equations} We will analyze Landau's equations for this diagram by anticipating the way we will use Feynman parameters to solve eq. (\ref{radjet2}). For the integration in $z$ we will first combine  the first and second denominator and subsequently the resulting denominator with the last denominator in eq. (\ref{radjet2}). Landau's equations for the resulting denominator for the $z$ integration read
\begin{align}
&z^2-2z\cdot (\alpha_2(\alpha_1y+(1-\alpha_1)w)+(1-\alpha_2)x)+\alpha_2(\alpha_1y^2+(1-\alpha_1)w^2)+(1-\alpha_2)x^2=0\label{land12}\\
&z^\mu=\alpha_2(\alpha_1y^\mu+(1-\alpha_1)w^\mu)+(1-\alpha_2)x^\mu\label{land22}
\end{align}
Let us analyze the case when $\boxed{\alpha_2=1}$, which corresponds to the case of a \textbf{soft emitted gluon} (since the Feynman parameter of the external gluon line is $1-\alpha_2$). In this case eq. (\ref{land22}) amounts to $z^\mu=\alpha_1y^\mu+(1-\alpha_1)w^\mu$ so that the emitted gluon does not change the direction of the fermion traveling from $w$ to $y$ as we expect from the momentum space picture. Using this result in eq. (\ref{land12}) for lightlike external points, $x^2=y^2=0$, one obtains  $\alpha_1(1-\alpha_1)(w-y)^2=0$, which sets $w-y$ on the lightcone except for the endpoints of the integration in $\alpha_1$, these endpoints correspond to UV divergences (in two distinct fermion lines) $z^\mu=w^\mu$ for $\alpha_1=0$ and $z^\mu=y^\mu$ for $\alpha_1=1$. Next, let us take $\boxed{\alpha_2=0}$, this gives the UV condition that $x^\mu=z^\mu$ so that the emitted gluon travels no distance. For general $\alpha_2$ and $\alpha_1=0,1$ we will have that $z$ will lie in the line connecting $x$ to $w$ and $y$ respectively.

Landau's equations for the other vertex, after the steps taken to reach eq. (\ref{intermediatestep2}) below, read
\begin{align}
\alpha w^2+2w\cdot \theta+\kappa^2&=0\;,\\
\alpha w^\mu+\theta^\mu&=0\;.
\end{align}
Here $\alpha=\alpha_4\alpha_5+(1-\alpha_1)\alpha_2(1-(1-\alpha_1)\alpha_2)$, $ \theta^\mu=-(1-\alpha_4)\alpha_5n^\mu-\alpha_1\alpha_2(1-\alpha_1)y^\mu-(1-\alpha_1)(1-\alpha_2)y^\mu$, and $\kappa^2=\alpha_1\alpha_2(1-\alpha_1\alpha_2)y^2+\alpha_2(1-\alpha_2)x^2-2\alpha_1\alpha_2(1-\alpha_2)x\cdot y$. Taking lightlike and collinear external points, one finds similar conclusions as in the previous example by following the same logic. 

\paragraph{Computation} As in the previous example we will use sequential Feynman parameters to perform the integration in $z$. We will first combine the first and second denominator and the resulting denominator with the last one in eq. (\ref{radjet2}), after shifting $z\to z-(\alpha_2(\alpha_1 y+(1-\alpha_1)w)+(1-\alpha_2)x)$, dropping odd powers and integrating in $z$ yields
\begin{align}
J^{(1)\mu b}_{\text{ext, rad}}(x,y)&=C^b
\int d^dw \int_0^\infty d t   \int_0^1d\alpha_1d\alpha_2\frac{\alpha_1^{1-\epsilon}(1-\alpha_1)^{1-\epsilon}\alpha_2^{3-2\epsilon}(1-\alpha_2)^{-\epsilon}}{(w^2+i\eta)^{2-\epsilon}((w-tn)^2+i\eta)^{1-\epsilon}} \Big(K^\mu-\gamma^\mu\slashed n  \slashed w(m^2)/2\Big)(m^2)^{2\epsilon-3} \;.\label{radjet22}
\end{align}
where $C^b$ is a color constant, $m^2=\alpha_1\alpha_2(1-\alpha_1\alpha_2)y^2+(1-\alpha_1)\alpha_2(1-(1-\alpha_1)\alpha_2)w^2+\alpha_2(1-\alpha_2)x^2-2(\alpha_1\alpha_2 y\cdot((1-\alpha_1)w+(1-\alpha_2)x)+(1-\alpha_1)(1-\alpha_2)w\cdot x)$, and $K^\mu=(\alpha_2(\alpha_1  \slashed y+(1-\alpha_1) \slashed w)+ \slashed y)\gamma^\mu (\alpha_2(\alpha_1  \slashed y+(1-\alpha_1) \slashed w)+ \slashed w) \slashed n\slashed w$. Next we combine the first and second denominators in eq. (\ref{radjet22}) and compute the integral in $t$ in the spacelike region $w\cdot n<0$. We combine the resulting two denominators with the last denominator left in two steps to get 
\begin{align}
J^{(1)\mu b}_{\text{ext, rad}}(x,y)&=C''^b\int d^dw \int_0^1d\alpha_1...d\alpha_5{\alpha_1^{1-\epsilon}(1-\alpha_1)^{1-\epsilon}\alpha_2^{3-2\epsilon}(1-\alpha_2)^{-\epsilon}\alpha_3^{1-\epsilon}(1-\alpha_3)^{-1-\epsilon}}\times\nonumber\\
&\times \frac{\alpha_4^{1-2\epsilon}\alpha_5^{2-2\epsilon}(1-\alpha_5)^{2-2\epsilon}}{(\alpha w^2+2w\cdot  \theta+\kappa^2)^{6-4\epsilon}}\Big(K^\mu-\gamma^\mu\slashed n  \slashed w(m^2)/2\Big)\;.\label{intermediatestep2}
\end{align}
We are now ready to compute the integral in $w$ by shifting $w\to w\alpha^{\frac{1}{2}}+ \theta\alpha^{-\frac{1}{2}}$ to obtain
\begin{align}
J^{(1)\mu b}_{\text{rad}}(x,y)&=\frac{ig^3T^b C_F}{2^{15-4\epsilon}\pi^{14-7\epsilon}}\int_0^1d\alpha_1...d\alpha_5\alpha_1^{1-\epsilon}(1-\alpha_1)^{1-\epsilon}\alpha_2^{3-2\epsilon}(1-\alpha_2)^{-\epsilon}\alpha_3^{-1-\epsilon}(1-\alpha_3)^{2-2\epsilon} \alpha_4^{2-3\epsilon}(1-\alpha_4)^{1-\epsilon}\times\nonumber\\
&\times\alpha_5^{4-4\epsilon}\alpha^{-2+\epsilon}\Big(\alpha g^{\rho\sigma}A^\mu_{\rho\sigma}\Gamma(3-3\epsilon)(M^2)/2+B^\mu\Gamma(4-3\epsilon)\Big)(M^2)^{3\epsilon-4}\;.\label{eq:23222}
\end{align}
The expressions for the $A_{\rho\sigma}^\mu$, $B^\mu$, and $M^2$ are shown next (here $\beta^\mu\equiv\alpha^{-\frac{1}{2}}\theta^\mu$):
\begin{align}
M^2=\kappa^2-\beta^2\;,
\end{align}
\begin{align}
B^\mu=&\Big[-(\alpha_2(\alpha_1  \slashed y+(\alpha_1-1) \slashed \beta)+ \slashed y)\gamma^\mu (\alpha_2(\alpha_1  \slashed y+(\alpha_1-1) \slashed \beta)- \slashed \beta)+\gamma^\mu \big(\alpha_1\alpha_2(1-\alpha_1\alpha_2)y^2\nonumber+\\
&+(1-\alpha_1)\alpha_2(1-(1-\alpha_1)\alpha_2)\beta^2+\alpha_2(1-\alpha_2)x^2-2[\alpha_1\alpha_2y\cdot((1-\alpha_2)x-\beta)-(1-\alpha_1)(1-\alpha_2)x\cdot \beta]\big)/2\Big]\slashed n\slashed \beta\;,
\end{align}
\begin{align}
A_{\rho\sigma}^\mu=&(\alpha_2(\alpha_1  \slashed y+(\alpha_1-1) \slashed \beta)+ \slashed y)\gamma^\mu(1+\alpha_2(1-\alpha_1))\gamma_\rho\slashed n \gamma_\sigma+(1-\alpha_1)\alpha_2\gamma_\rho \gamma^\mu\big[(\alpha_2(\alpha_1  \slashed y+(\alpha_1-1) \slashed \beta)- \slashed \beta)\slashed n \gamma_\sigma-\nonumber\\
&-(1+\alpha_2(1-\alpha_1))\gamma_\sigma \slashed n\slashed \beta \big]+\gamma^\mu\slashed n \big[(1-\alpha_1)\alpha_2(1-(1-\alpha_1)\alpha_2)\slashed \beta g_{\rho\sigma}/2+\gamma_\rho \big((1-\alpha_1)\alpha_2(1-(1-\alpha_1)\alpha_2)\beta_\sigma+\nonumber\\
&+\alpha_1\alpha_2(1-\alpha_1)y_\sigma +(1-\alpha_1)(1-\alpha_2)x_\sigma \big)\big]\;.
\end{align}
The result in eq. (\ref{eq:23222}) has the same form as eq. (\ref{eq:23}) and could be combined to produce a more tractable result, although the inclusion of the other two diagrams in the abelian case is needed for gauge invariance. We will leave this for a future work.
For completeness, we now turn to analyze Landau's equations for the self energy corrections to the one-loop radiative jet function.

\subsubsection{Landau's equations for the internal-emission self-energy correction to the one-loop radiative jet function}

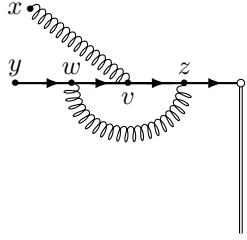
\begin{figure}[ht!]
\begin{tikzpicture}[>=stealth]
\draw (1.5,-0.2) node {$v$};
\draw (2.25,0.2) node {$z$};
\draw (.75,0.2) node {$w$};
 \draw [decoration={aspect=0.4, segment length=1mm, amplitude=1mm,coil},decorate] (0.75,0) .. controls (0.75,-0.9) and (2.25,-.9) .. (2.25,0);
  \draw [decoration={aspect=0.4, segment length=1mm, amplitude=1mm,coil},decorate] (.2,1) -- (1.5,0);
  \draw[-latex,thick] (2.6,0)--(2.75,0);
  \draw[-latex,thick] (1,0)--(1.25,0);
    \draw[-latex,thick] (1.75,0)--(2,0);
  \draw[-latex,thick] (0.35,0)--(.6,0);
    \draw [thick](0,0)--(2.95,0);
  \draw [thick](1,0)--(2.1,0);
   \draw (0.75,-0.00) node {\tiny$\bullet$};
   \draw (2.25,-0.00) node {\tiny$\bullet$};
   \draw (1.5,-0.00) node {\tiny$\bullet$};
      \draw (.2,0.99) node {\tiny$\bullet$};
      \draw (0,-0.00) node {\tiny$\bullet$};
   \draw[double] (3,-0.05)--(3,-2);
   \draw (2.95,0) arc [start angle=180, end angle=-180, radius=0.05cm];
      \draw (0,1) node {$x$};
          \draw (0,0.2) node {$y$};
\end{tikzpicture}
\centering
\caption{Diagram representing the one-loop self energy with internal gluon emission from a jet.}\label{repera3}
\end{figure}
Next, I will analyze Landau's equations for the self-energy contributions to gain some intuition on some of their divergent configurations. The first one is depicted in Fig. \ref{repera3} and, after Feynman parametrization, equals 

\begin{align}
J^{(1)\mu b}_{\text{SE, int, rad}}(x,y)&=(ig^3T^aT^bT^a)\frac{\Gamma(10-6\epsilon)}{(2\pi^{2-\epsilon})^4(4\pi^{2-\epsilon})^2}
\int d^dzd^dw\int_0^1d\alpha'_1... d\alpha'_6{\alpha'_1}^{1-\epsilon}{\alpha'_2}^{1-\epsilon}{\alpha'_3}^{1-\epsilon}{\alpha'_4}^{-\epsilon}{\alpha'_5}^{1-\epsilon}{\alpha'_6}^{-\epsilon} \times\nonumber\\
&\times\delta\left(1-\sum_{n=1}^6 \alpha'_n\right) \frac{(\slashed y-\slashed w)\gamma^\nu(\slashed w-\slashed v)\gamma^\mu(\slashed v-\slashed z)\gamma_\nu \slashed z}{\big(\alpha'_1(y-w)^2+\alpha'_2(w-v)^2+\alpha'_3(v-z)^2+\alpha'_4(w-z)^2+\alpha'_5z^2+\alpha'_6(x-v)^2\big)^{10-6\epsilon}}\;.\label{radjet3}
\end{align}

First of Landau's equations for the diagram in Fig. \ref{repera3} reads
\begin{align}
&&\alpha'_1(y-w)^2+\alpha'_2(w-v)^2+\alpha'_3(v-z)^2+\alpha'_4(w-z)^2+\alpha'_5z^2+\alpha'_6(x-v)^2=0\label{landauselfenergy1}\;,
\end{align}
which as usual sets all lines on the lightcone except if the lines are hard ($z_j\to0$) or soft ($\alpha_j=0$).  

Now, we will treat the integration in each vertex separately, effectively imposing conservation of momentum by employing Feynman parameters in separate steps (just as in the previous examples). The primed Feynman parameters in eq. (\ref{landauselfenergy1}) will be replaced later with the corresponding parameters after successive Feynman parametrizations. Furthermore, we will set all lines not connected to the vertex under analysis to be lightlike. 

For the integration in $z$, the second Landau's equation gives (taking $\alpha_3'=\alpha_3(1-\alpha_4)$, $\alpha_5'=(1-\alpha_3)(1-\alpha_4)$ and $\alpha_4'=\alpha_4$)
\begin{align}
&z^\mu={\alpha_3(1-\alpha_4) v^\mu+\alpha_4 w^\mu}\;.
\end{align}
If either the emitted gluon or the emitted fermion at $z$ is soft ($\alpha'_3=0$ and $\alpha'_4=0$ respectively) then $z$ must lie in the direction of $w$ or $v$ respectively. So that the emitted particle does not change the direction of the particle emitting it. \\
If $\alpha'_5=1$ and hence $\alpha_3=\alpha_4=0$ (soft emitted lines from $z$) then there is a UV solution to Landau's equations with $z^\mu=0$. \\
For $\alpha'_5=0$ with $\alpha_3=1$ (so that the outgoing fermion from the Wilson line cusp is soft), $z^\mu$ will be the convex combination of the other two internal vertices (\textit{i.e.} on the line connecting them) and yields the condition that $(1-\alpha_4)\alpha_4(v-w)^2=0$. So that if $\alpha_4\neq 0,1$, then $v$ and $w$ must be lightlike separated in order to obtain a divergence in the amplitude. 
For the case when the emitted gluon is soft, $\alpha_4=0$, then the gluon does not change the direction of the fermion, $z^\mu=\alpha_3 v^\mu$, and $v^2=0$ except for the UV cases where $\alpha_3=0$ ($z=0$) and $\alpha_3=1$ ($z=v$). If $\alpha_3=0$ then it is the outgoing fermion from $z$ that is soft and hence $z^\mu=\alpha_4 w^\mu$ and hence $w^2=0$ except for the UV cases.\\

For the integration in $v$, the second Landau's equation gives (taking $\alpha_2'=\alpha_2(1-\alpha_6)$, $\alpha_3'=(1-\alpha_2)(1-\alpha_6)$ and $\alpha_6'=\alpha_6$)
\begin{align}
&v^\mu={\alpha_2(1-\alpha_6) w^\mu+(1-\alpha_2)(1-\alpha_6) z^\mu+\alpha_6 x^\mu}\label{eq:vint}\;.
\end{align}
Let us analyze the case when the emitted gluon traveling to $x$ becomes soft, \textit{i.e.} ${\alpha_6=0}$. In this case $v^\mu$ will lie on the line connecting the other two internal vertices not changing the direction of the fermion. Introducing eq. (\ref{eq:vint}) into eq. (\ref{landauselfenergy1}) we find $\alpha_2(1-\alpha_2)(w-z)^2=0$. This means that, if no fermion line connected to $v$ is soft, then the fermion lines connecting $z$ to $w$ must lie on the lightcone. If $\alpha_2$ or $(1-\alpha_2)$ are equal to zero then the fermion lines can be off the lightcone and there is a UV singularity since we will have that  $v=z$ or $v=w$ respectively. \\
The integration in $w$ has very similar Landau's equations as the one in $z$. Hence, we will move on to analyze these equations for the next and last contribution to the abelian one-loop radiative jet function.

\subsubsection{Landau's equations for the external-emission self-energy correction to the one-loop radiative jet function}

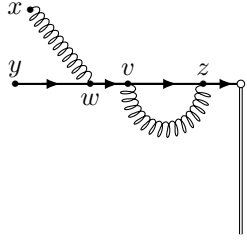
\begin{figure}[ht!]
\begin{tikzpicture}[>=stealth]
      \draw (0,1) node {$x$};
          \draw (0,0.2) node {$y$};
 \draw (1.5,.2) node {$v$};
 \draw (2.5,.2) node {$z$};
  \draw (1,-.2) node {$w$};
 \draw [decoration={aspect=0.4, segment length=1mm, amplitude=1mm,coil},decorate] (1.5,0) .. controls (1.5,-0.8) and (2.5,-.8) .. (2.5,0);
  \draw [decoration={aspect=0.4, segment length=1mm, amplitude=1mm,coil},decorate] (.2,1) -- (1,0);
  \draw[-latex,thick] (2.,0)--(2.15,0);
    \draw[-latex,thick] (2.5,0)--(2.9,0);
 \draw[-latex,thick] (1.,0)--(1.375,0);
  \draw[-latex,thick] (0.35,0)--(.6,0);
    \draw [thick](0,0)--(2.95,0);
  \draw [thick](1,0)--(2.1,0);
   \draw (1,-0.00) node {\tiny$\bullet$};
   \draw (1.5,-0.00) node {\tiny$\bullet$};
    \draw (2.5,-0.00) node {\tiny$\bullet$};
      \draw (.2,0.99) node {\tiny$\bullet$};
      \draw (0,-0.00) node {\tiny$\bullet$};
   \draw[double] (3,-0.05)--(3,-2);
   \draw (2.95,0) arc [start angle=180, end angle=-180, radius=0.05cm];
\end{tikzpicture}
\centering
\caption{Diagram representing the one-loop self energy with external gluon emission from a jet.}\label{repera4}
\end{figure}

The last contribution to the abelian one-loop radiative jet function is shown in Fig. \ref{repera4} and, after Feynman parametrization, amounts to

\begin{align}
J^{(1)\mu b}_{\text{SE, ext, rad}}(x,y)&=(ig^3T^bC_F)\frac{\Gamma(10-6\epsilon)}{(2\pi^{2-\epsilon})^4(4\pi^{2-\epsilon})^2}
\int d^dzd^dw\int_0^1d\alpha'_1... d\alpha'_6{\alpha'_1}^{1-\epsilon}{\alpha'_2}^{1-\epsilon}{\alpha'_3}^{1-\epsilon}{\alpha'_4}^{-\epsilon}{\alpha'_5}^{1-\epsilon}{\alpha'_6}^{-\epsilon} \times\nonumber\\
&\times\delta\left(1-\sum_{n=1}^6 \alpha'_n\right) \frac{(\slashed w-\slashed y) \gamma^\mu(\slashed w-\slashed v)(d-2)(\slashed v-\slashed{z})\slashed{z}}{\big(\alpha'_1(y-w)^2+\alpha'_2(w-v)^2+(\alpha'_3+\alpha'_4)(v-z)^2+\alpha'_5z^2+\alpha'_6(x-w)^2\big)^{10-6\epsilon}}\;.\label{radjet4}
\end{align}

Hence, first of Landau's equations for the diagram in Fig. \ref{repera4} reads
\begin{align}
&&\alpha'_1(y-w)^2+\alpha'_2(w-v)^2+(\alpha'_3+\alpha'_4)(v-z)^2+\alpha'_5z^2+\alpha'_6(x-w)^2=0\;,\label{landauselfenergy2}
\end{align}
which again sets all lines on the lightcone except if the lines are hard ($z_j\to0$) or soft ($\alpha_j=0$). Notice that for the self energy piece $\alpha_3+\alpha_4$ serves the purpose of being one Feynman parameter only.\\

For the integration in $w$, the second Landau's equation gives (taking $\alpha_1'=\alpha_1(1-\alpha_6)$, $\alpha_2'=(1-\alpha_1)(1-\alpha_6)$ and $\alpha_6'=\alpha_6$)
\begin{align}
&w^\mu={\alpha_1(1-\alpha_6) y^\mu+(1-\alpha_1)(1-\alpha_6) v^\mu+\alpha_6 x^\mu}\label{eq:wint}\;.
\end{align}
Giving the usual solution for the soft-emitted gluon case ${\alpha_6=0}$, $w$ lying in the line connecting $v$ to $y$ and $\alpha_1(1-\alpha_1) (y-v)^2=0$. So that the fermion lines connecting $y$ and $v$ must lie on the lightcone and the soft gluon does not change the direction of the fermion. Again, UV solutions arise whenever $\alpha_1=0,1$ with $w=v$ and $w=y$ respectively. An analogue situation happens for a soft emitted fermion from $w$ to $y$ ($v$) corresponding to $\alpha_1=0$ ($\alpha_1=1$) for general $\alpha_6$. 
To finish let us analyze Landau's equations for one vertex in the self energy part, for example $v$ (the one with $z$ is very similar). The second Landau's equation gives (taking $\alpha_3'=\alpha_3(1-\alpha_2)$, $\alpha_4'=(1-\alpha_3)(1-\alpha_6)$ and $\alpha_2'=\alpha_2$)
\begin{equation}
v^\mu=\alpha_2 w^\mu+(1-\alpha_2)z^\mu\;,
\end{equation}
so that $v$ must lie on the straight line connecting $w$ and $z$ with the usual condition $\alpha_2(1-\alpha_2)(w-z)^2=0$.

\newpage 
 \section{Conclusion}
In this paper I have discussed the results leading to factorization of QCD amplitudes in both momentum and coordinate spaces and fully solved the one-loop contribution to the jet function in coordinate space, finding that the most divergent part of it is proportional to a fermion propagator. This fact was identified as a configuration where the gluon emerging from the cusp of the Wilson line travelled collinearly with the fermion to the external point, producing hence the divergence. An LSZ reduction of my result shows that we recover the correct leading divergence expected from the previously known momentum space results.

In view of the growing interest on NLP divergences of QCD factorized amplitudes, I have also reduced two radiative contributions to the one-loop jet function into a Feynman parameters integral. Finally, I have analyzed Landau's equations for all contributions (without external leg corrections) to the abelian one-loop radiative jet function, highlighting the relevance of these equations in revealing the divergent (soft and collinear) configurations in the amplitudes. I have also studied how separating divergent terms in Feynman parameters has an interesting potential in factorizing different pinched regions as well as making explicit their overlapping.\\ 
\section{Acknowledgments}

I want to thank my supervisors Eric Laenen and Felipe J. Llanes-Estrada for excellent guidance and the NIKHEF staff for creating a welcoming atmosphere. Also, I want express my gratitude to the financial support of the Amsterdam Science Talent Scholarship, the Spanish Grant MICINN PID2019-108655GB-I00 and the IPARCOS institute.


\end{document}